\documentclass[fleqn,10pt]{wlscirep}
\usepackage[utf8]{inputenc}
\usepackage[T1]{fontenc}
\usepackage{amssymb}
\usepackage{subfig}
\usepackage{graphicx}
\usepackage{geometry}
\usepackage{array}
\usepackage{multirow}

\usepackage{orcidlink}

\usepackage{hyperref}

\title{Uncertainty Quantification Using Ensemble Learning and Monte Carlo Sampling for Performance Prediction and Monitoring in Cell Culture Processes}

\author[1,*]{Thanh Tung Khuat \orcidlink{0000-0002-6456-8530}}
\author[2]{Robert Bassett}
\author[2]{Ellen Otte}
\author[1]{Bogdan Gabrys \orcidlink{0000-0002-0790-2846}}
\affil[1]{Complex Adaptive Systems Laboratory, The Data Science Institute, University of Technology Sydney, NSW 2007, Australia}
\affil[2]{CSL Innovation, Melbourne, VIC 3000, Australia}

\affil[*]{Corresponding author: thanhtung.khuat@uts.edu.au}

\begin{abstract}
Biopharmaceutical products, particularly monoclonal antibodies (mAbs), have gained prominence in the pharmaceutical market due to their high specificity and efficacy. As these products are projected to constitute a substantial portion of global pharmaceutical sales, the application of machine learning models in mAb development and manufacturing is gaining momentum. This paper addresses the critical need for uncertainty quantification in machine learning predictions, particularly in scenarios with limited training data. Leveraging ensemble learning and Monte Carlo simulations, our proposed method generates additional input samples to enhance the robustness of the model in small training datasets. We evaluate the efficacy of our approach through two case studies: predicting antibody concentrations in advance and real-time monitoring of glucose concentrations during bioreactor runs using Raman spectra data. Our findings demonstrate the effectiveness of the proposed method in estimating the uncertainty levels associated with process performance predictions and facilitating real-time decision-making in biopharmaceutical manufacturing. This contribution not only introduces a novel approach for uncertainty quantification but also provides insights into overcoming challenges posed by small training datasets in bioprocess development. The evaluation demonstrates the effectiveness of our method in addressing key challenges related to uncertainty estimation within upstream cell cultivation, illustrating its potential impact on enhancing process control and product quality in the dynamic field of biopharmaceuticals.
\end{abstract}
\begin{document}

\flushbottom
\maketitle
%
%
\thispagestyle{empty}

\section{Introduction}
In recent years, biopharmaceutical products, including monoclonal antibodies (mAbs) and therapeutic proteins derived from biological organisms for the treatment or prevention of diseases, have emerged as top-selling drugs in the pharmaceutical market \cite{luhw20}. This trend is attributed to their numerous advantages, such as high specificity and activity \cite{homo23}. As global pharmaceutical sales are projected to surpass \$1 trillion by 2026, biopharmaceutical products are expected to contribute significantly, constituting 37\% of the total sales, an increase from 30\% in 2020 \cite{ph21}. By 2026, over half of the top 100 best-selling medications are anticipated to be biologics. Within the realm of biological products, monoclonal antibodies (mAbs) stand out as the forefront runners in the swiftly expanding market of high-value biologics \cite{pabu19}. The mAb products are manufactured through biotechnological processes within living systems, including microorganisms, plants, animals, or human cells such as Chinese hamster ovary (CHO) cells, mouse myeloma (NS0), baby hamster kidney (BHK), human embryo kidney (HEK-293), and human retinal cells \cite{wu04}. Hence, the cultivation and harvesting of cells responsible for producing the active pharmaceutical ingredient \cite{paqu17} play a crucial role in facilitating the growth and reproduction of cells in quantities sufficient to meet production demands \cite{sa22}. Continuing this trajectory, the utilization of machine learning models across different phases of monoclonal antibody (mAb) development and manufacturing, including the prediction and monitoring of biophysical properties, cell growth, nutrient, metabolite, and protein concentrations throughout bioreactor cell cultivation processes, is not only gaining popularity but also improving in accuracy \cite{khba24}.

According to Kelley \cite{ke09}, the attention in cell culture process development has been shifted from solely pursuing the elevated titers to emphasising the control of product quality and process consistency at every stage of the development and across all production scales. Therefore, it is crucial to monitor the changes in the culture operating parameters which include physical, chemical and biological parameters. Physical parameters encompass factors such as temperature, gas flow rate, and agitation speed, while chemical parameters encompass dissolved oxygen and carbon dioxide levels, pH, osmolality, nutrient  and metabolite concentrations. Biological parameters are employed to assess the physiological condition of the culture and include metrics such as viable cell concentration, viability, and a range of intracellular and extracellular measurements \cite{livi10}. Optimizing culture operating parameters is essential to attain high product expression while maintaining acceptable product quality profiles. This purpose can be achieved by monitoring the relationships among process variables and extracting valuable knowledge from bioprocess data using machine learning models to gain novel insights into the interdependence between Critical Process Parameters (CPPs) and Critical Quality Attributes (CQAs) in biopharmaceutical process development and manufacturing. To construct bioprocess datasets and calibrate the machine learning models, it is essential to measure process parameters during the cell culture process within bioreactors. These process parameters can be measured either online or offline with operator intervention. Examples of offline measurements include pH (often for verification of online pH readings), cell counting, and viability measurements, osmolality, and specific metabolite and product concentrations. Metabolites in cell culture, such as glucose, lactate, glutamine, and glutamate, are typically assessed offline using enzymatic biosensors designed for each specific analyte \cite{livi10}. These measurements play a crucial role not only in sustaining substrate levels above critical thresholds through feeding strategies but also in formulating processes with minimised by-product formation.

Commercially available auto-samplers and integrated multi-functional offline analysers, such as the BioProfile FLEX, have typically been used for offline monitoring of metabolite levels, osmolality, pH, dissolved gases, and measurements of sodium, potassium, and calcium. This is done as a replacement for manual sampling, which is often labor-intensive and can introduce operator-dependent errors into the process \cite{deab10}. Although autosamplers and analysers can obtain a high accuracy, analyses in \cite{mcmc12} showed that the coefficient of variation among different measurement times ranges from 3\% to 8\% for each process parameter. This fact reflects the uncertainty of input features and target variables when building predictive models. To assist in making control decisions based on the predictive outcomes of machine learning models, it is necessary to provide uncertainty levels associated with each predictive value. In the context of the regression problem addressed in this paper, the variance in predictive outcomes for an input query can serve as a meaningful indicator of uncertainty. A comprehensive review paper conducted by Hullermeier and Waegeman \cite{homo23} categorised uncertainty sources into aleatoric (data-dependent, noise-induced) and epistemic (model-dependent) uncertainties. Aleatoric uncertainty, often termed as the irreducible component of uncertainty, is associated with randomness, which is the variability in the outcome of an experiment due to inherently random effects. This type of uncertainty cannot be mitigated through model enhancements. Instead, reducing aleatoric uncertainty is achievable through improvements in the data, such as incorporating repeat measurements or eliminating erroneous entries \cite{hemc23}. On the contrary, epistemic uncertainty represents the reducible uncertainty stemming from insufficient knowledge about the optimal model, and it can be diminished through model enhancements \cite{huvo21}. This type of uncertainty can be further divided into uncertainties arising from the selection of the model (including architecture, representations, and features) and the ambiguity in parameter optimisation once a model is selected.

Numerous methods for quantifying uncertainty in predictive outcomes of machine learning models are available in the literature, as outlined in \cite{huvo21}. Among them, the two most popular groups are ensemble learning and Bayesian methods \cite{khba24}. While Bayesian methods such as Gaussian Process regression focus mainly on quantification and reducing the epistemic uncertainty, ensemble methods aim to estimate the impact of the  aleatoric uncertainty due to the use of sampling techniques on the input data. 

This paper focuses on introducing a general framework for uncertainty quantification applicable to any regressors, especially in the context of small training data, using ensemble learning. In situations involving limited training data and the presence of noise in input features, the conventional approach of developing multiple base learners through bootstrap resampling from input spaces in ensemble learning becomes inefficient. To estimate the impact of the aleatoric uncertainty on the prediction accuracy in small training datasets, we propose the use of Monte Carlo simulations to generate additional input samples by considering available training features as mean values. The additional instances will be used to train base learners. The effectiveness of the proposed method will be assessed in the context of predicting and monitoring the performance of process parameters during upstream cell culture bioreactor runs. One of the inherent challenges in cell culture processes pertains to the limitation of available data, commonly referred to as the small data issue. This limitation arises from the scarcity of process data for emerging bioproducts, with instances where only one or two production runs are conducted for a novel product at manufacturing sites \cite{tuga18, baal21}. The substantial cost associated with cell culture processes further exacerbates this issue, as conducting bioreactor runs for new cell lines or experimental variations (e.g., novel base media or feeding strategies) is economically constrained. Additionally, the practical necessity of relocating products across different production sites to accommodate various products and their life cycles contributes to what is known as the small training data problem. This operational characteristics result in a limited number of historical experiments available at new manufacturing facilities. The adoption of Process Analytical Technology (PAT) tools in bioprocess development and manufacturing steps has facilitated the real-time collection of extensive and diverse measurements and information. This can lead to the availability of thousands of input features, for example, each Raman spectrum from the spectrometer can contain thousands of spectrum variables (e.g., wave numbers) considered as input features. Meanwhile, the number of experiments (i.e., samples) is limited. Consequently, these circumstances give rise to a Low-N problem, wherein the number of training samples is considerably smaller than the number of input data dimensions. This inherent disparity poses a considerable challenge for machine learning algorithms, when the number of training samples is inadequate relative to the number of input features.

By generating new values for input features and the target variable based on their actual values, along with a coefficient of variation associated with each input feature, we can overcome the shortage issue of training data when training an ensemble model of multiple base learners. In our proposed method, we will use the standard deviation value of the predictive outcomes of all base learners as an indicator of the uncertainty level for each predicted value. In short, our novel contribution can be summarized as follows:
\begin{enumerate}
    \item We introduce a comprehensive framework for assessing the uncertainty level linked to each predictive value through the utilization of ensemble learning of regressors in tandem with Monte Carlo sampling. Our proposed method is designed to address the challenge of limited training data, a factor that can affect the effectiveness of traditional ensemble learning approaches. Moreover, our method represents the general framework employing ensemble learning in conjunction with Monte Carlo simulations to quantify the uncertainty level associated with each predictive outcome, particularly in scenarios characterised by a shortage of training data.
    \item The effectiveness of the proposed method is evaluated through its application to two prominent challenges in upstream cell cultivation within bioreactors. The first problem involves the early prediction of antibody concentrations one day in advance, utilising solely current offline measurements as input features. The second problem entails real-time monitoring of glucose concentrations throughout the bioreactor run of a cell culture process, employing Raman spectra as input features. 
\end{enumerate}

The subsequent sections of this paper are organised as follows. Section \ref{method} provides an introduction to the key features of the proposed method. In Section \ref{offline_performance}, the efficacy of the proposed approach is evaluated in addressing the challenge of predicting process performance in cell culture bioreactors one day in advance, using solely offline process measurements of the bioreactors. Section \ref{monitoring} demonstrates the effectiveness of the proposed method in addressing the real-time monitoring problem of glucose concentrations during bioreactor runs, using Raman spectra data as input features. Finally, the concluding remarks of this paper will be presented in Section \ref{conclusion}.

\section{Methodology} \label{method}
Because of the errors linked with offline measurements, as discussed in the Introduction section, relying solely on a single predicted value generated by machine learning models for each set of input features proves inadequate for making informed decisions during the cell culture process. This limitation becomes evident, for instance, in scenarios where it may not suffice to determine crucial actions such as deciding when to add glucose or terminate the cell cultivation process. It becomes challenging to assess the accuracy of a predicted value without considering the associated uncertainty range. With predictions that include uncertainty values, we gain knowledge of possible minimum and maximum values associated with each prediction. Hence, it is preferable to have results presented in the form of $\hat{y} \pm 2 \cdot \sigma $, where $\hat{y}$ represents the predicted value, and $\sigma$ signifies the standard deviation of the prediction. The prediction values may not follow a normal distribution, so $2 \cdot \sigma$ will be used to derive confidence limits, providing approximately 95\% certainty that the actually observed values will fall within the prediction range. This section outlines a method to achieve this objective by employing an ensemble of regressors and Monte Carlo sampling on both input and output spaces to construct training sets.

Let $\mathbf{X} = [X_1, X2, \ldots, X_m]$ be a set of $m$ input samples, where each sample $X_k = (x_{k1}, x_{k1}, \ldots, x_{kn})$ ($k \in [1, m]$) includes $n$ input features, and $\mathbf{Y} = [Y_1, Y2, \ldots, Y_m]$ ($Y_k = \{y_k\}, y_k \in \mathbb{R}, k \in [1, m])$ be outputs corresponding to input samples. We need to build a regressor $\mathbb{F}(\mathbf{X}) \rightarrow \hat{\mathbf{Y}} $ such that minimises $||\mathbf{Y} - \mathbf{\hat{Y}}||$ value. The output of the regressor $\mathbb{F}$ for each unseen input sample $X_T$ will be in the form of $\hat{y}_T \pm \sigma_T$. In this case, $\hat{y}_T$ is the average predictive value of $N$ base learners within the ensemble model computed by Eq.~\eqref{mean_pred}, while $\sigma_T$ represents the standard deviation value of $N$ predictive values given in Eq.~\eqref{std_pred}.
\begin{equation}\label{mean_pred}
\hat{y}(X_T) = \cfrac{1}{N} \cdot \sum_{i = 1}^N{\hat{y}_i(X_T)}
\end{equation}
where $\hat{y}_i(X_T)$ is the predicted value of the $i^{th}$ base regressor for an unseen input sample $X_T$ within the ensemble model, which comprises $N$ base regressors. The standard deviation of the ensemble prediction for each data point ($X_T$) can be employed to establish confidence intervals and uncertainty bounds. The utilised standard deviation is the unbiased standard deviation derived from the predictions of an individual base model for each data point:
\begin{equation}\label{std_pred}
    \sigma(X_T) = \sqrt{\cfrac{\sum_{i=1}^N{(\hat{y}_i(X_T) - \hat{y}(X_T))^2}}{N - 1}}
\end{equation}

\begin{figure}[!ht]
    \centering
    \includegraphics[width=\textwidth]{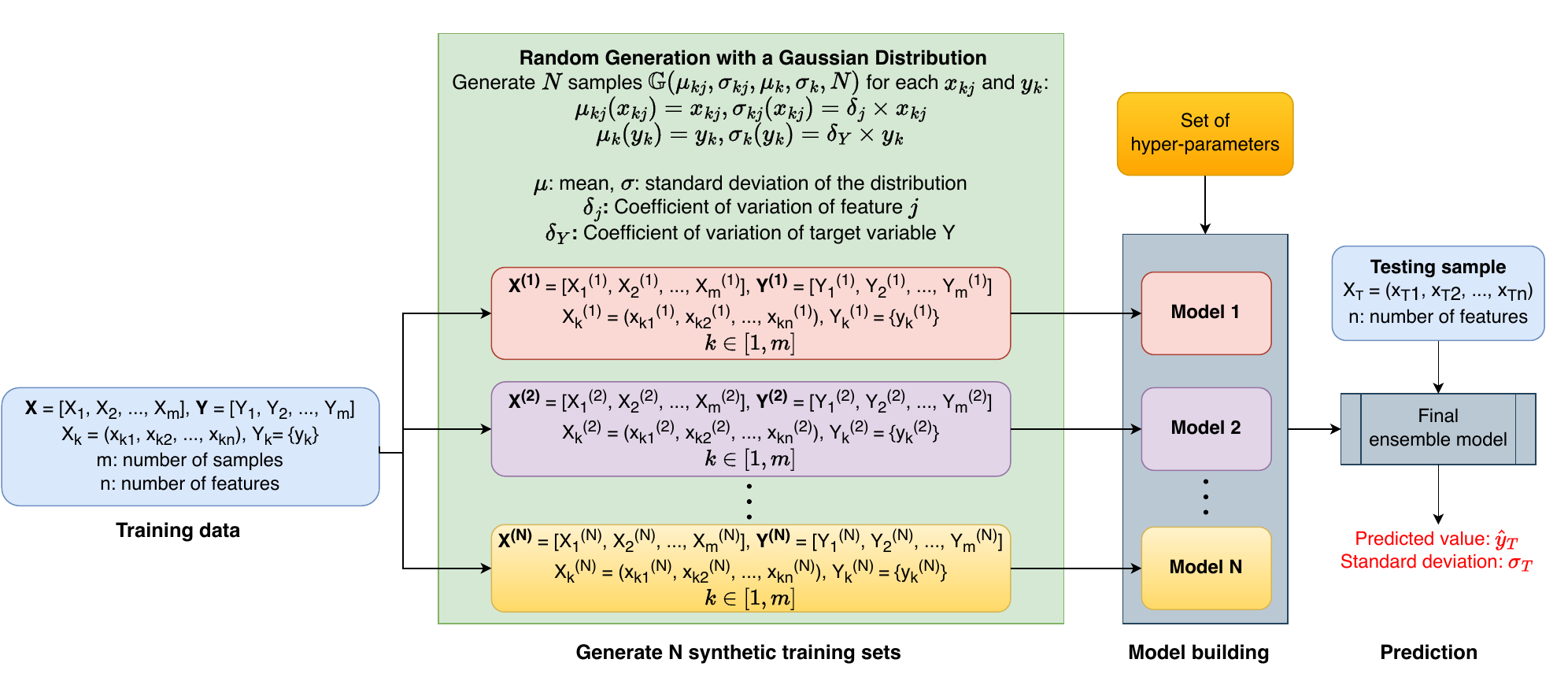}
    \caption{General framework for estimating uncertainty levels of predictive values using ensemble learning and Monte Carlo sampling.}
    \label{framework}
\end{figure}

To train $N$ base regressors of an ensemble model, we need to generate $N$ training sets, each for training a base regressor. We will not use resampling or random sampling without replacement from original training set to create $N$ training sub-sets because the training data set is small in size, so the traditional sampling techniques may not generate sufficient diversity in the training sub-sets for the base regressors within the ensemble model. In addition, each input value is subjected to the errors due to the variation in measurements resulting from the final accuracy of offline analysers. As a result, in our proposed method, the Monte Carlo sampling method is employed to generate $N$ random values for each input feature $x_{kj}$ ($k \in [1, m], j \in [1, n]$) and each target value $y_k$ satisfying a Gaussian distribution requirement with mean being an actual value ($x_{kj}$ or $y_k$) and given standard deviation value. In a mathematical form, we generate $N$ random values from a Gaussian distribution $\{x_{kj}^{(1)}, 
\ldots, x_{kj}^{(N)}\} = \mathbb{G}(\mu_{kj}=x_{kj}, \sigma_{kj}, N)$ for each input feature $x_{kj}$ and $\{y_{k}^{(1)}, 
\ldots, y_{k}^{(N)}\} = \mathbb{G}(\mu_{k}=y_k, \sigma_{k}, N)$ for each target value $y_k$, where $\sigma_{kj}$ and $\sigma_k$  are the standard deviations for each input feature $x_{kj}$ and target value $y_k$ respectively. These standard deviation values are computed from corresponding actual values and coefficient of variation ($\delta$) of each input feature as follow: $\sigma_{kj} = \delta_j \times x_{kj}$ and $\sigma_{k} = \delta_Y \times y_k$, where $\delta_j$ is the maximum coefficient of variation of feature $j$, while $\delta_Y$ is the maximum coefficient of variation of output variable $\mathbf{Y}$. After generating $N$ samples for all input and output values, we will concatenate all values $x_{kj}^{(i)}$ and $y_{k}^{(i)}$ at the $i^{th}$ position ($i \in [1, N]$) to create the $i^{th}$ training set $\mathbf{X}^{(i)}$ and $\mathbf{Y}^{(i)}$ in order to train the $i^{th}$ base regressor within the ensemble model. The fundamental steps of the proposed framework are presented in Fig. \ref{framework}.

To optimise the hyper-parameters of the ensemble models including the number of base regressors $N$ and the hyper-parameters of base regressors, we need a separate validation set and find the set of hyper-parameters resulting in the minimum average value of errors over $N_{val}$ samples in the validation sets. For instance, we can seek for a given set of hyper-parameters to obtain the minimum mean absolute error (MAE) of the ensemble model with $N$ base regressors:
\begin{equation}
    MAE = \cfrac{\sum_{k=1}^{N_{val}}(|y(X_k) - \cfrac{1}{N} \cdot \sum_{i=1}^N{\hat{y}_i(X_k)}|)}{N_{val}}
\end{equation}
where $y(X_k)$ is the target value of the $k^{th}$ sample in the validation set, and $\hat{y}_i(X_k)$ is the prediction of the $i^{th}$ base regressor in the ensemble model for a validation sample $X_k$. Although our proposed framework can be used for any regression models as base learners, this work only illustrates the empirical outcomes for two types of regression models. The first model is the support vector regression, which is one of the most often used machine learning algorithms for small datasets \cite{koko22}. The second model is the Partial Least Squares Regression (PLSR), which is very popular for small data \cite{gole12} and dominates in machine learning models applied for predicting a variety of process performance issues in cell culture processes \cite{khba24}.

\section{Performance prediction of a cell culture process} \label{offline_performance}
\subsection{Introduction to the problem}
Chinese hamster ovary (CHO) cells are commonly employed for the production of monoclonal antibodies. The cell culture procedure involves a sequence of scale-up and expansion stages designed to yield a sufficient cell mass for the inoculation of the production bioreactors. This process includes additional cell growth, monoclonal antibody production, the elimination of cellular mass from the bioreactor material through centrifugation and a three-stage filtration process resulting in the acquisition of clarified material \cite{homo23}. In the course of biopharmaceutical process development, it is crucial to enhance titer aiming to reduce manufacturing expenses but still maintaining consistent quality attributes, safety and efficacy of therapeutic proteins. Throughout this development phase, continuously enhancing upstream titer will play a critical role in increasing the output and reducing the costs of production \cite{xure18}. The optimisation of nutrient concentrations, including amino acids, vitamins, and trace metals, is widely recognised as a crucial factor for enhancing the protein production \cite{kile05}. In addition, the cell concentration and viability play a pivotal role in the development of cell culture processes. These measurements are essential for assessing the culture physiology in response to operating conditions, calculating growth rates, specific consumption/production rates of metabolites, and determining cell-specific productivity \cite{livi10}. Therefore, there is a high expectation of making early predictions of mAb concentration in the upcoming culture days based on current values of tens of offline measurements, such as operational conditions, nutrients, and metabolite concentrations monitored over time. Accurate predictions will contribute to adjusting the culture environments and nutrient compositions to increase the product concentration. This section will consider the effectiveness of the proposed framework in predicting mAb concentration one-day-ahead using values of offline measurements at the current day.

\subsection{Dataset}
The dataset utilized in this study was derived from AstraZeneca's upstream process development and production databases, encompassing various antibody products employing CHO cell lines, as detailed in \cite{gase21}. The dataset included information from 106 cultures, reflecting a diverse operational scale ranging from bench-top (5L volume) to manufacturing (500L volume), spanning a period of 7 years from 2010 to 2016. Each culture involved the recording of over 20 offline parameters for up to 17 days, including Culture Days, Elapsed Culture Time (ECT), Viable cell density (VCD), Total cell density (TCD), pH, Cell Viability, Elapsed Generation Number (EGN), Average Cell Volume (ACV), Osmolality, Average Cell Compactness (ACC), Average Cell Diameter (ACD),  Cumulative Population Doubling Level (CPDL), concentrations of glucose, lactate, ammonium, glutamine, glutamate, sodium, potassium, and bicarbonate, as well as temperature, pCO2, pO2, monomer content of the final product, and product (mAb) concentration. The time-series dataset has been normalised to safeguard proprietary rights.

We have used all offline measurements in the original dataset as input features and created a new target variable, which is the mAb concentration of the next culture day. We have also removed the sample corresponding to the last culture day as there is no value in the target variable.

\subsection{Learning procedures}
In this experiment, we employed two types of regression models, namely the partial least squares regression (PLSR) and the support vector regression (SVR), as base learners within the ensemble model for small-sized datasets shown in Figure \ref{framework}. The dataset, extracted from \cite{gase21}, lacks information regarding the coefficient of variation for each offline measurement. Consequently, a fixed value of $\delta_j = \delta_Y = 0.05$ was used for all offline measurements serving as input features and the target variable. In addition to the ensemble models of PLSR and SVR, we conducted separate training for PLSR, SVR, and Gaussian Process (GP) Regression as competing models for performance comparison. The GP regression has notable advantages such as lower data requirements and more importantly the possibility to assess the uncertainty of predictions \cite{huvo21}. The implementation of PLSR, SVR, and GP was taken from the scikit-learn library \cite{peva11}.

We will assess the obtained performance of all learning models based on errors through a 5-fold group cross-validation. The set of 106 cultures will be partitioned into five folds, with each fold encompassing data from all culture days within a specific culture. Four folds will serve for training, while the remaining fold will be designated as testing data. This process will be iterated five times, with each bioreactor used once in a testing fold. The average error of the trained models across the five testing folds will be employed for comparing the performance among competing models. For each training fold, we employed a hyper-parameter optimisation procedure using the Bayesian optimization approach within the Optuna library \cite{aksa19}. This optimization involved 50 iterations and 5-fold cross-validation to determine the optimal hyperparameter settings before training the model on the respective training fold. The potential range of hyper-parameter values for each regression model is detailed in Table \ref{opt_range}. It is noted that all individual regressors within the ensemble model will use the same hyper-parameter setting.

\begin{table}
\centering
\renewcommand{\arraystretch}{0.1}
\begin{tabular}{|l|m{12cm}|}
\hline
\textbf{Model} & \textbf{Range} \\
\hline
Ensemble of SVR models & \begin{itemize}[leftmargin=*, itemsep=0pt]
    \item Number of base regressors ($N$): [30, 200]
    \item Regularization parameter ($C$): [$10^{-4}$, $10^5$] in the logarithm domain
    \item Kernel coefficient for `rbf' kernel ($\gamma$): [$10^{-4}$, $10^2$] in the logarithm domain
    \item Epsilon parameter ($\epsilon$): [$10^{-3}$, $10^2$] in the logarithm domain
\end{itemize} \\
\hline
Ensemble of PLSR models & \begin{itemize}[leftmargin=*, itemsep=0pt]
    \item Number of base regressors ($N$): [30, 200]
    \item Number of components ($n\_components$) [2, 20]
\end{itemize}  \\
\hline
SVR & \begin{itemize}[leftmargin=*, itemsep=0pt]
    \item Regularization parameter ($C$): [$10^{-4}$, $10^5$] in the logarithm domain
    \item Kernel coefficient for `rbf' kernel ($\gamma$): [$10^{-4}$, $10^2$] in the logarithm domain
    \item Epsilon parameter ($\epsilon$): [$10^{-3}$, $10^2$] in the logarithm domain
\end{itemize} \\
\hline
PLSR & \begin{itemize}[leftmargin=*, itemsep=0pt]
    \item Number of components ($n\_components$): [2, 20]
\end{itemize}  \\
\hline
GP & \begin{itemize}[leftmargin=*, itemsep=0pt]
    \item Kernel: \{DotProduct, Matern, RBF, RationalQuadratic \}
    \item The length scale or inhomogenity of the kernel: [$10^{-3}$, $10^3$] in the logarithm domain
    \item Variance of additional Gaussian measurement noise ($\alpha$): [$10^{-3}$, $10^{-1}$] in the logarithm domain
\end{itemize}  \\
\hline
\end{tabular}
\caption{The search range of hyper-parameters for machine learning models.} \label{opt_range}
\end{table}

\subsection{Evaluation metrics}
The public dataset given in Gangadharan et al. \cite{gase21} was normalised to the range of 0 to 1. With the existence of values of 0, several metrics with the actual values in the denominator such as Mean Absolute Percentage Error (MAPE) will not work. In this experiment, we will use Mean Absolute Error (MAE) as a performance metric to compare the learning models:
\begin{equation}
    MAE = \cfrac{\sum_{i=1}^{N_{test}}{|\hat{y}_i - y_i|}}{N_{test}}
\end{equation}
where $N_{test}$ is the number of testing samples, $\hat{y}_i$ is the prediction of the $i^{th}$ testing sample, and $y_i$ is the true value of the $i^{th}$ testing sample.

For the learning models which return the standard deviation associated with the predictive values, we will compute both MAE scores for the upper bound ($\hat{y} + 2 \sigma$) and the lower bound ($\hat{y} - 2 \sigma$) as follows:
\begin{equation}
    MAE^+ = \cfrac{\sum_{i=1}^{N_{test}}{|(\hat{y}_i + 2\sigma_i) - y_i|}}{N_{test}}
\end{equation}
\begin{equation}
    MAE^- = \cfrac{\sum_{i=1}^{N_{test}}{|(\hat{y}_i - 2\sigma_i) - y_i|}}{N_{test}}
\end{equation}
where $\sigma_i$ is the standard deviation associated with the prediction $\hat{y}_i$ of the $i^{th}$ testing sample. If the $MAE$ value is small, while the values of $MAE^+$ and $MAE^-$ are high, the average of predictions provided by all individual base learners contributes to the reduction of variations among individual learners. This case also indicates that the uncertainty level in the predictive outcomes is high. When the level of uncertainty is high, decision-making based on the prediction results should be approached with caution.

\subsection{Empirical results and discussions}
\begin{figure}[!ht]
\centering
\includegraphics[width=0.6\textwidth]{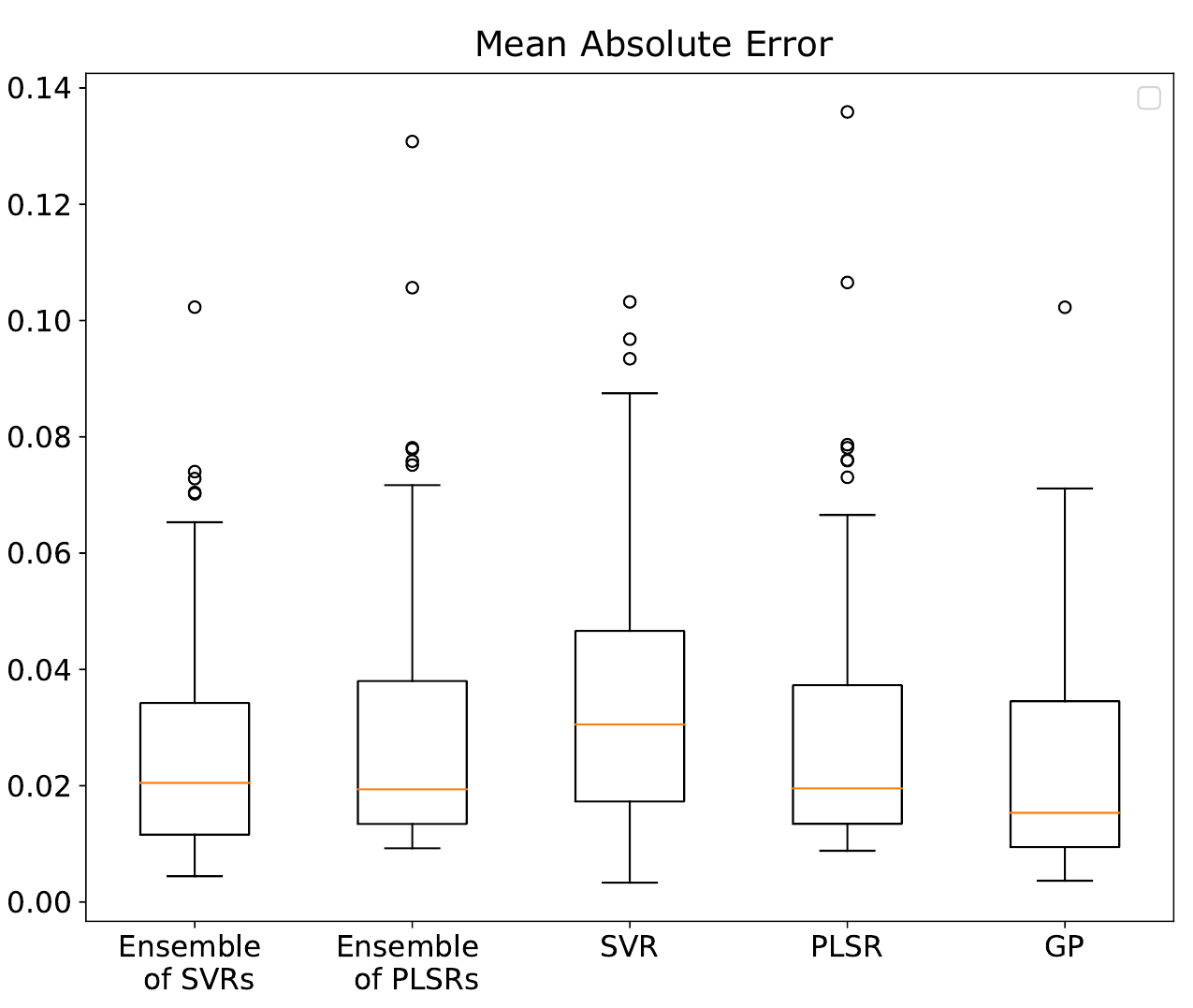}
\caption{Comparing the performance of different machine learning models in predictions of all 106 bioreactors used in the testing set over 5-fold group cross-validation.}\label{mab_all_bio}
\end{figure}

This section compares the average performance of the proposed ensemble models with individual models such as SVR, PLSR, and GP. Table \ref{result_mab} summarises the mean MAE scores and standard deviation values over 5-fold group cross-validation of predictions, upper bounds, and lower bounds. Meanwhile, Fig. \ref{mab_all_bio} shows box-and-whisker plots of the compared ML models for MAE scores of all 106 cultures used as testing data over 5-fold group cross-validation. It can be observed that the performance of the ensemble of SVRs outperforms that of using an individual SVR. However, the performance of the ensemble model of PLSRs is equal to the performance of a single PLSR model. When comparing the MAE scores of upper and lower bounds with the MAE score of predictive values generated by the average value of base regressors, we can see that the uncertainty level of predictions is small in this case. This is different from the case of using an individual GP model. In this experiment, although the GP can provide the best performance, the uncertainty level of predictions is higher than that using our proposed method. In addition, the predictive performance of all 106 cultures presented in Fig. \ref{mab_all_bio} illustrates that the ensemble of SVRs is competitive with the GP model.

\begin{table}
\centering
\renewcommand{\arraystretch}{1}
\begin{tabular}{|m{2cm}|c|c|c|c|c|}
\hline
 & \textbf{Ensemble of SVRs} & \textbf{Ensemble of PLSRs} & \textbf{SVR} & \textbf{PLSR} & \textbf{GP} \\
\hline
Prediction & 0.0256 $\pm$ 0.0042 & 0.0288 $\pm$ 0.0024 & 0.0344 $\pm$ 0.0106 & 0.0285 $\pm$ 0.0026 & 0.0233 $\pm$ 0.0035 \\
\hline
Upper bound & 0.0301 $\pm$ 0.0065 & 0.0293 $\pm$ 0.0024	& - & - & 0.0573 
$\pm$ 0.0033 \\
\hline
Lower bound	& 0.0287 $\pm$ 0.0036 & 0.0293 $\pm$ 0.0028	& -	& - & 0.0583 $\pm$ 0.0055\\
\hline
\end{tabular}
\caption{The mean and standard deviation of MAE scores over 5-fold group cross-validation for different ML models} \label{result_mab}
\end{table}

\begin{figure}[!ht]
\centering
\begin{subfloat}[Ensemble of SVR models (MAE: 0.0044)]{
\includegraphics[width=0.48\textwidth]{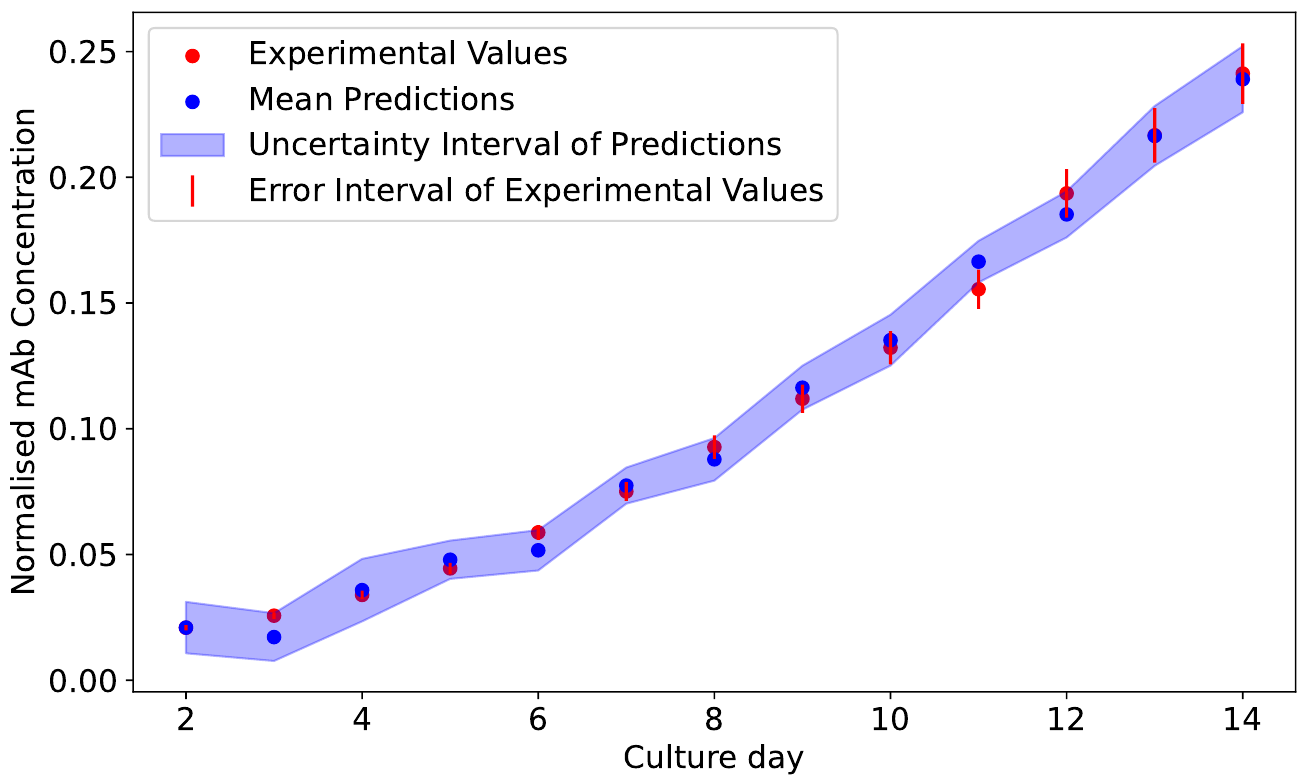}}
\end{subfloat}
\begin{subfloat}[Ensemble of PLSR models (MAE: 0.0092)]{
\includegraphics[width=0.48\textwidth]{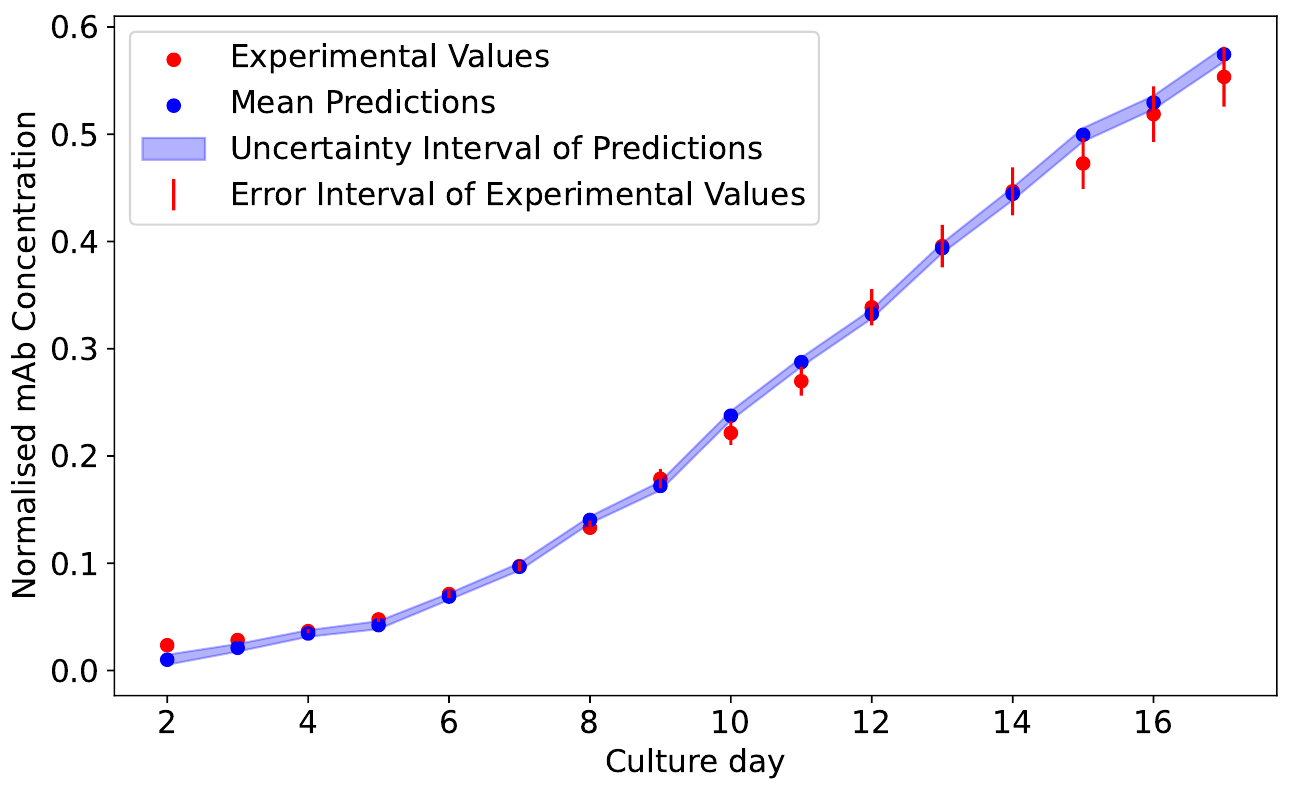}}
\end{subfloat}
\begin{subfloat}[SVR (MAE: 0.0033)]{
\includegraphics[width=0.48\textwidth]{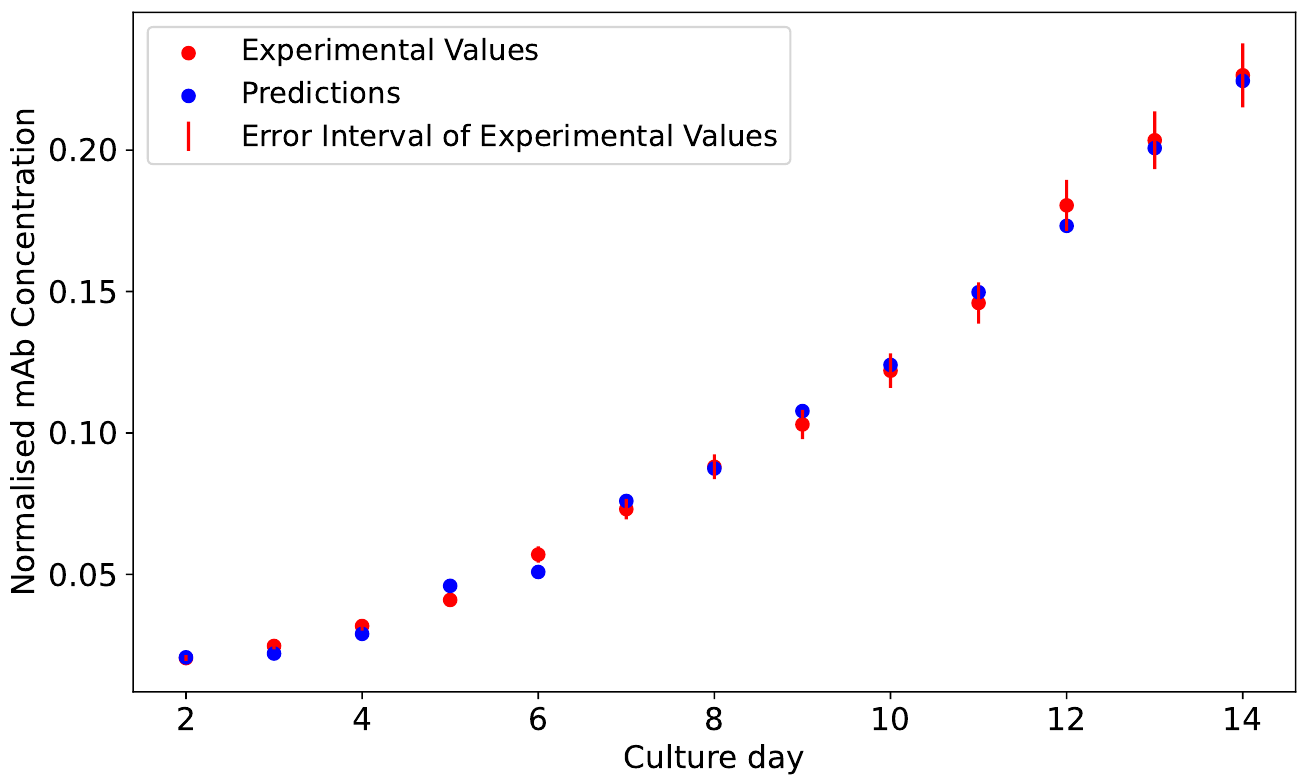}}
\end{subfloat}
\begin{subfloat}[PLSR (MAE: 0.0088)]{
\includegraphics[width=0.48\textwidth]{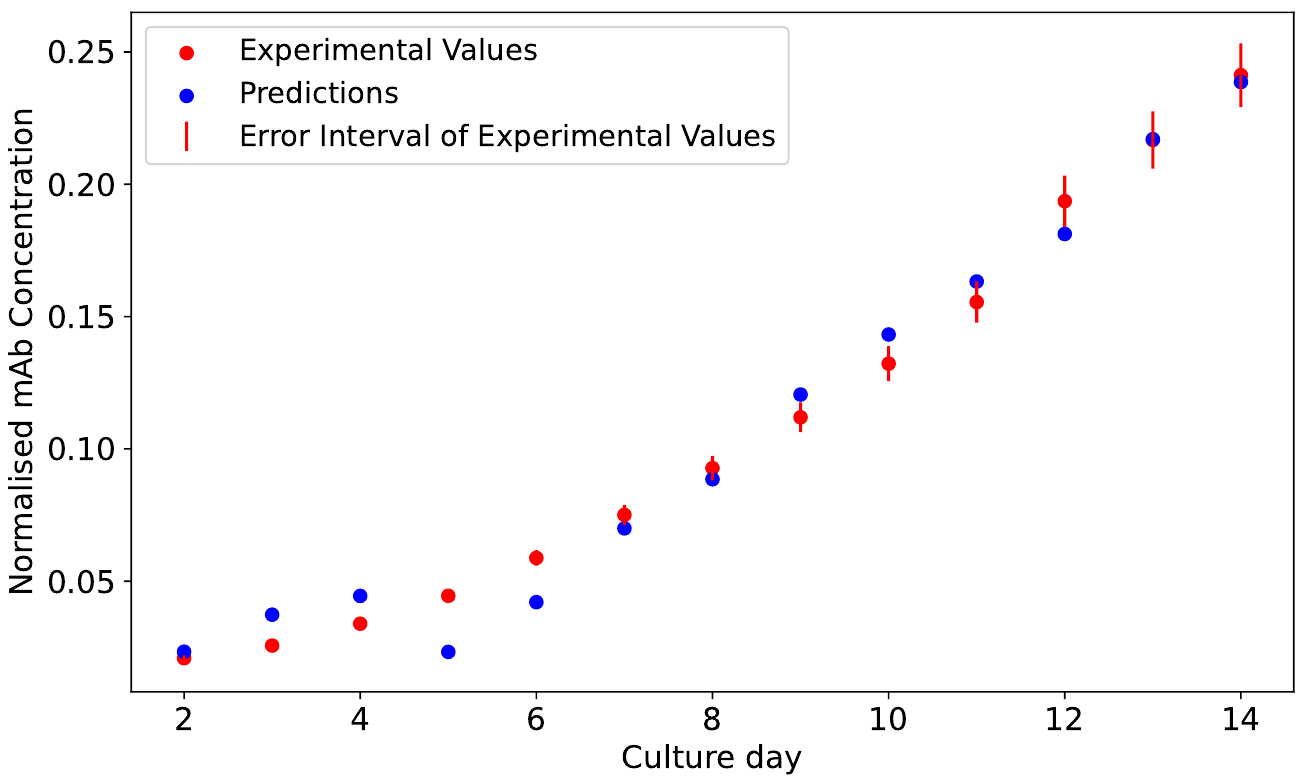}}
\end{subfloat}
\begin{subfloat}[GP (MAE: 0.0037)]{
\includegraphics[width=0.48\textwidth]{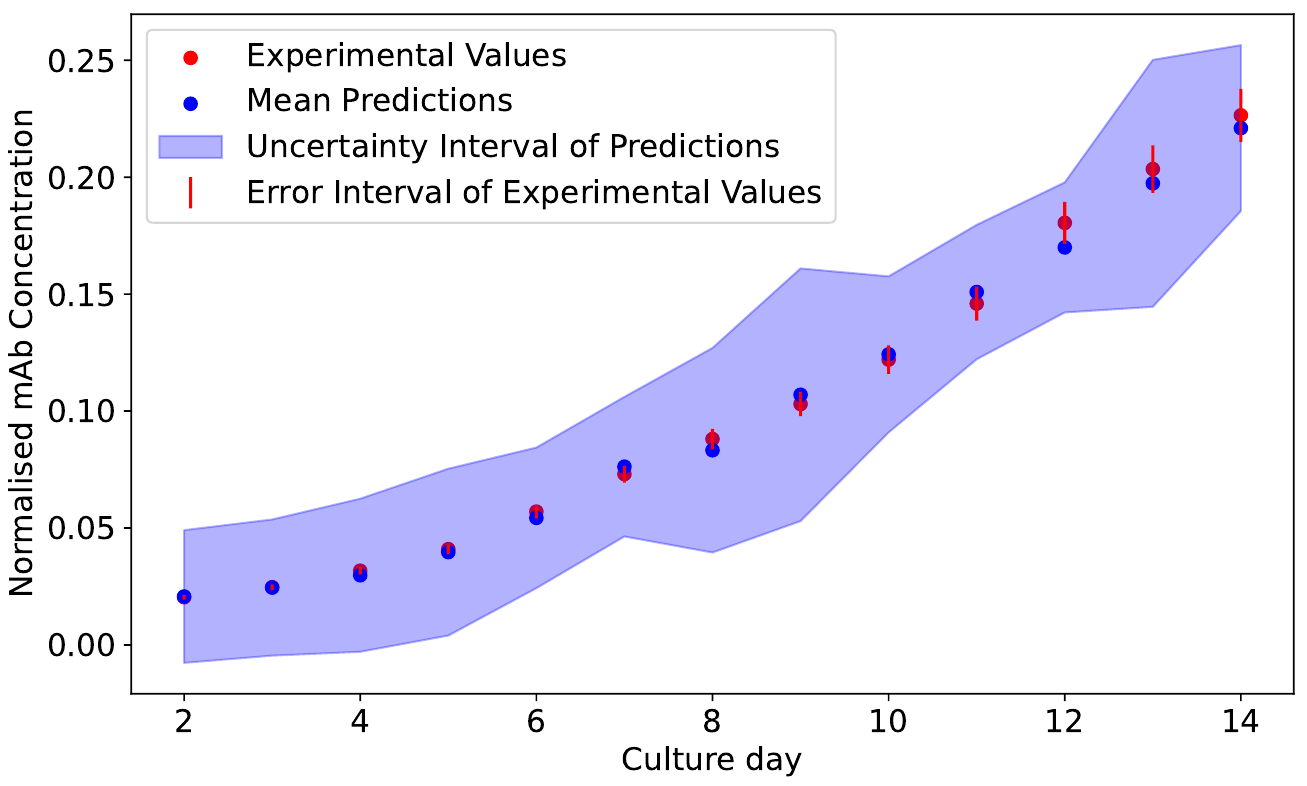}}
\end{subfloat}
\caption{The best prediction of each ML model.}\label{mab_best}
\end{figure}

\begin{figure}[!ht]
\centering
\begin{subfloat}[Ensemble of SVR models (MAE: 0.1023)]{
\includegraphics[width=0.48\textwidth]{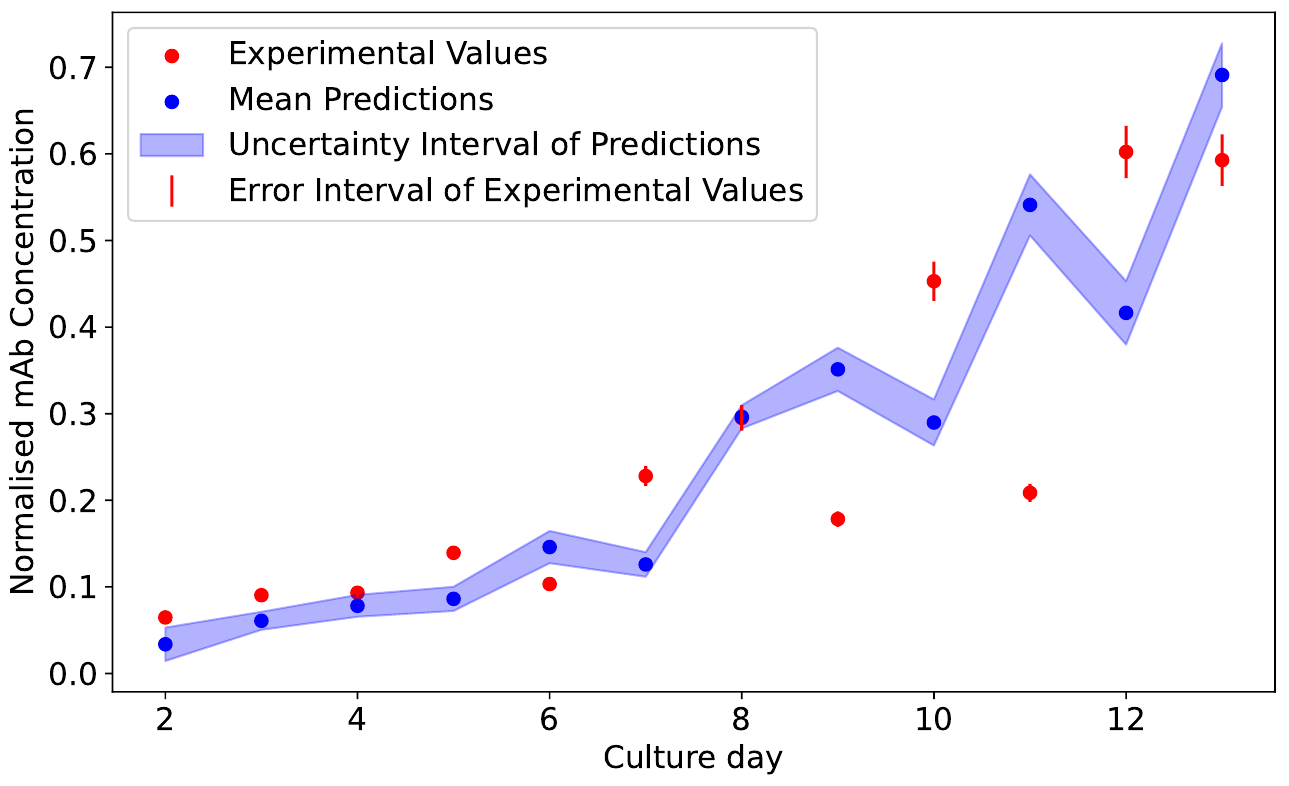}}
\end{subfloat}
\begin{subfloat}[Ensemble of PLSR models (MAE: 0.1308)]{
\includegraphics[width=0.48\textwidth]{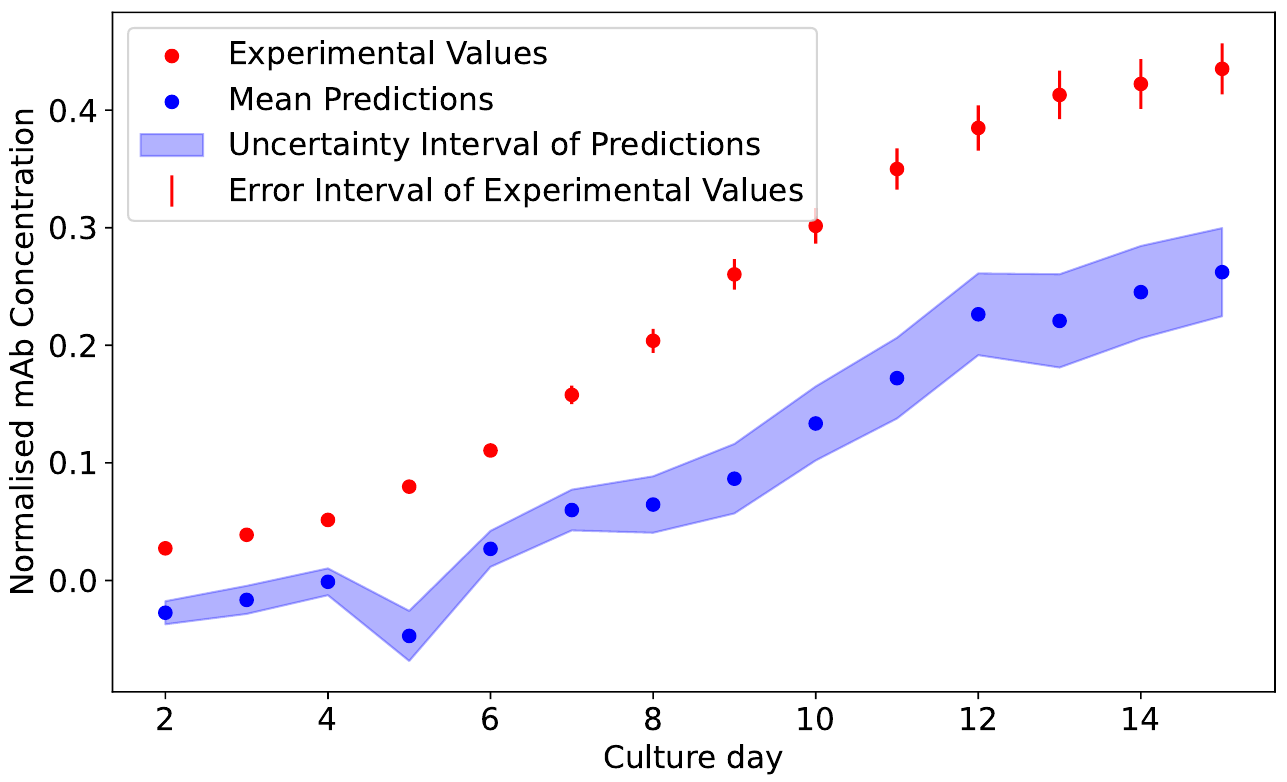}}
\end{subfloat}
\begin{subfloat}[SVR (MAE: 0.1032)]{
\includegraphics[width=0.48\textwidth]{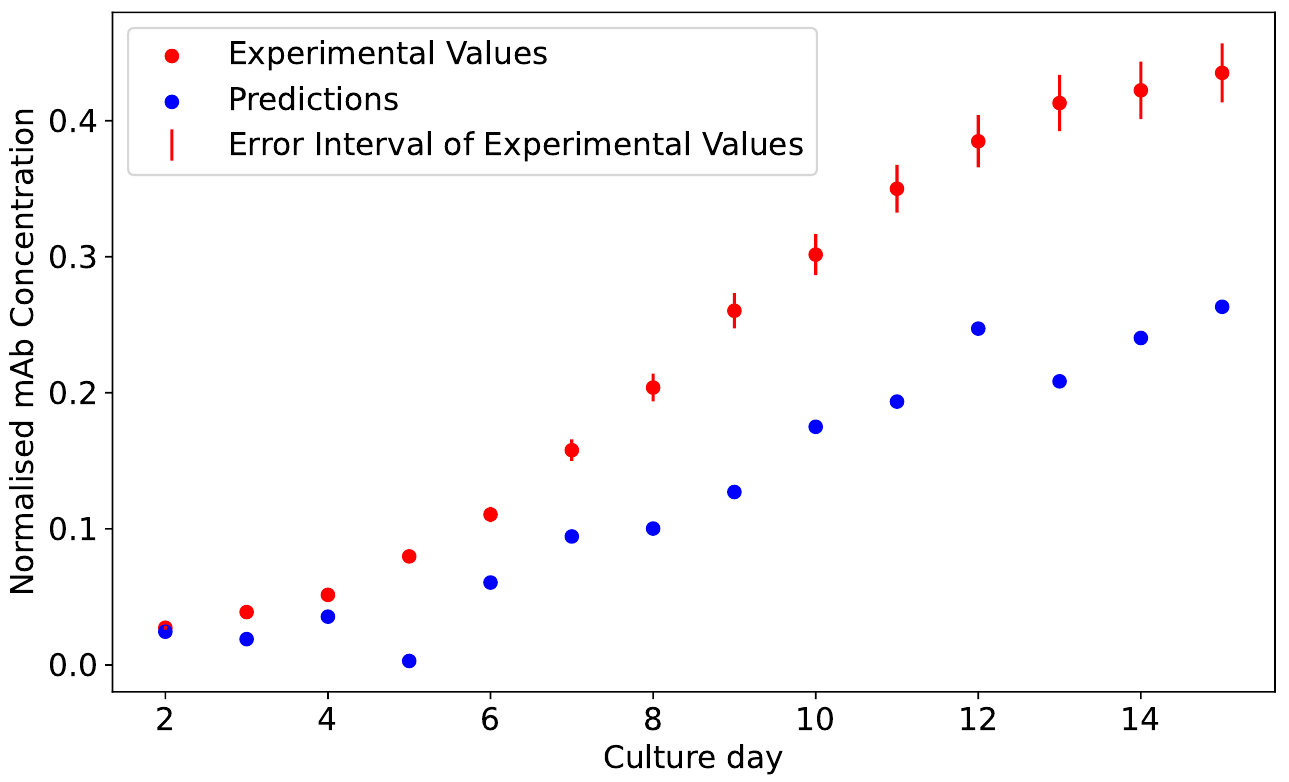}}
\end{subfloat}
\begin{subfloat}[PLSR (MAE: 0.1359)]{
\includegraphics[width=0.48\textwidth]{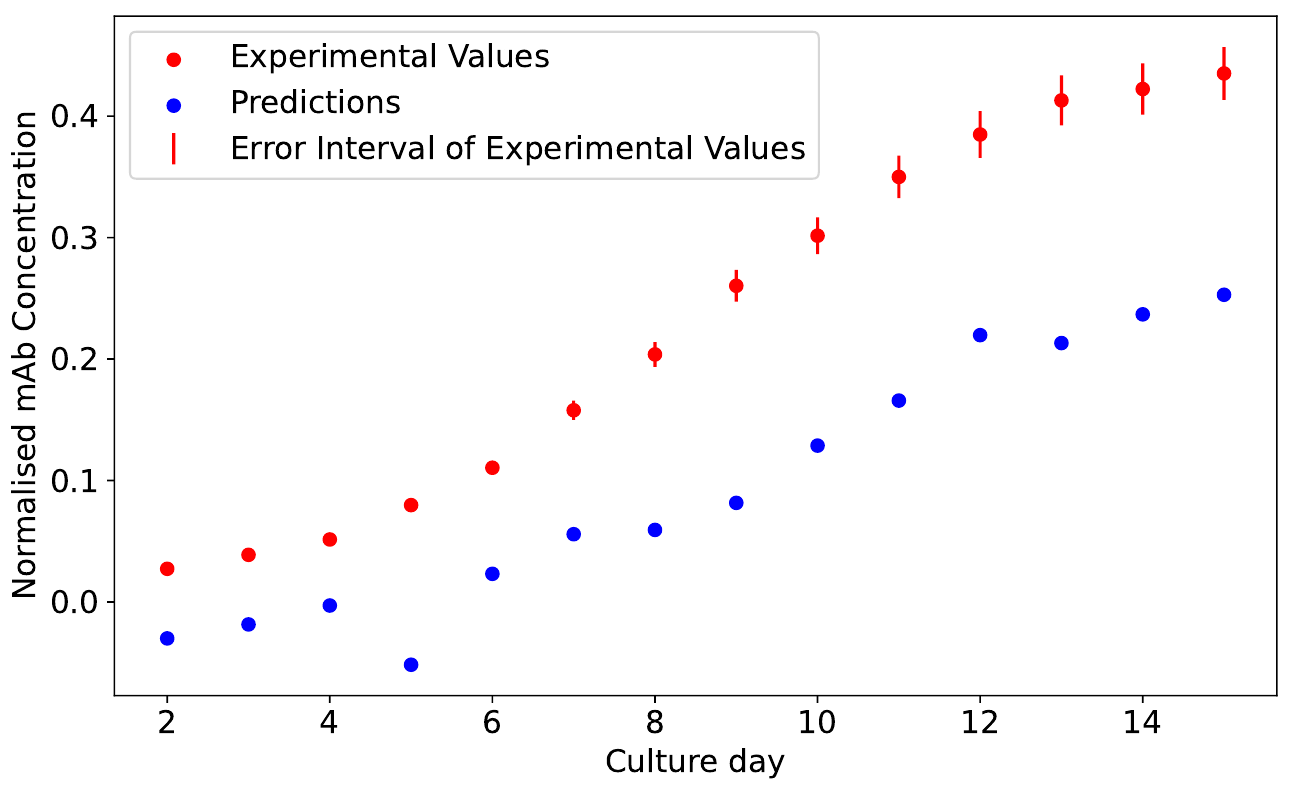}}
\end{subfloat}
\begin{subfloat}[GP (MAE: 0.1023)]{
\includegraphics[width=0.48\textwidth]{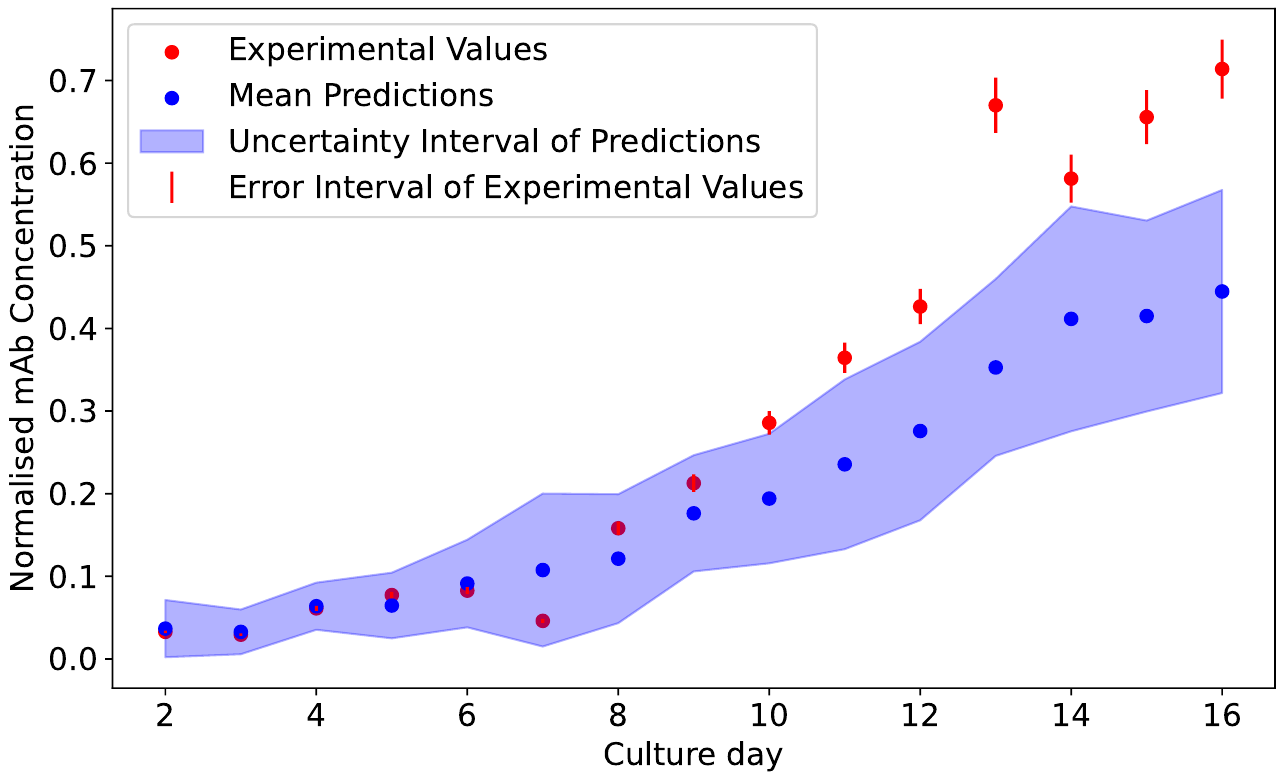}}
\end{subfloat}
\caption{The worst prediction of each ML model.}\label{mab_worst}
\end{figure}

Fig. \ref{mab_best} depicts the culture exhibiting the most accurate predictive performance among the 106 cultures using various ML models. In contrast, Fig. \ref{mab_worst} showcases the culture with the least accurate predictive performance. It is evident that when the mAb concentration gradually increases throughout the culture time, the predictive performance of ML models tends to be high. Conversely, when the mAb concentration fluctuates, either increasing or decreasing suddenly during the cell culture process, the performance of ML models typically diminishes, and the associated uncertainty of predicted values increases.

The outcomes presented in this section serve as a proof-of-concept for the proposed framework, operating under the assumption of a uniform coefficient of variation of 5\% for all offline measurements, given that all input features are normalised to the range of 0 and 1. In practical scenarios, offline measurements may possess varying coefficients of variation, impacting the predictive performance of ML models. When dealing with different coefficients of variation for distinct input features, it becomes crucial to assess the importance of each input feature in determining predictive outcomes.

In the upcoming section, we will address another scenario where real-time measured input features are presumed accurate without the presence of a coefficient of variation. However, the target variable, being a specific offline measurement, exhibits variations in accuracy across different measuring times.

\section{Real-time monitoring of glucose concentration within bioreactors using Raman spectra data} \label{monitoring}
\subsection{Introduction to the problem}
Currently, monitoring the cell culture profile during production involves taking small samples of medium components and metabolites at specific culture points, which are then quantified using a bioanalyser \cite{taki23}. However, this sampling process presents challenges, including potential effects on the culture volume and the risk of microbial contamination. Additionally, the limited number of sampling points makes it challenging to acquire data at high frequencies \cite{taki23}. As a result, various Process Analytical Technology (PAT) methods have been developed to enable continuous analysis \cite{khba24}. For instance, Raman spectrometers and near-infrared spectroscopy can offer information on components in the culture solution, while capacitance-based measurements allow for cellular concentration analysis \cite{giwa22}. The application of Raman spectrometers for the continuous acquisition of various culture data in cultivation processes enables real-time monitoring of cell growth, nutrient, and metabolite concentrations. This marks a crucial step towards implementing feedback controls for culture conditions. For instance, a Raman-based glucose feedback control mechanism can enhance overall bioreactor health, product output, and product quality \cite{gima23}. Moreover, real-time monitoring of culture components may expedite faster medium development by continuously optimising a broader range of components \cite{taki23}.

In this section, we will evaluate the efficacy of various ML models in real-time monitoring of glucose concentrations within bioreactors throughout the cell culture process, using only Raman spectra data as input features. The actual outputs of the target variable corresponding to input Raman spectra will be based on offline glucose measurements. Periodically sampled offline glucose concentrations will be analysed by bioanalysers such as Nova Biomedical BioProfile FLEX Analyser. As a result, the target variable will exhibit a coefficient of variation. Meanwhile, the input Raman data is high-dimensional, and the relationships and dependencies among input wave numbers (features) and between all input features and the output variable are complex and usually non-linear. Therefore, the value of the coefficient of variation for each input Raman feature is typically unknown. Consequently, this scenario differs from the issue discussed in Section \ref{offline_performance}, so it requires modifications to the framework outlined in Section \ref{framework}.

\subsection{Dataset}
In this experiment, we used a dataset extracted from the cell culture process within CSL Innovation Pty Ltd. The dataset includes the historical culture data of three 5L bioreactors (A1, A2, and A3) taking place in two weeks. All of the three bioreactors used the same culture media (base and feed) but A3 used a different cell line expressing a different product than A1 and A2, which were the same cell line and product.

Raman spectra were acquired within the bioreactor using a Kaiser Raman Rxn2 analyzer equipped with a 785 nm excitation laser and a probe. The spectra were collected in the Raman shift range of 100 to 3425.0 $cm^{-1}$. On average, approximately four spectra were recorded per hour during bioreactor runs. In contrast, glucose concentrations were sampled and analysed twice daily using the BioProfile FLEX Analyser. To enhance the accuracy of ML models, offline measurements were also taken before and immediately after glucose feeding. In total, there are 100 offline values of glucose concentrations for all three bioreactors over the two weeks of cell culturing.

Due to the mismatch of time points at which online Raman spectra and offline glucose concentration measurements were collected, it is necessary to map the Raman spectra to the corresponding offline glucose concentration values for building training and testing datasets. In this study, we will associate each offline glucose concentration value with the closest Raman spectra collected after the timestamp of the offline measurement. As there are no Raman spectra collected during the feeding process, the offline glucose concentration value acquired immediately before glucose feeding will be mapped to the closest Raman spectra collected prior to that specific offline collection timepoint.

\subsection{Data preprocessing}\label{raman_preprocessing}
\begin{figure}[!ht]
    \centering
    \includegraphics[width=0.6\textwidth]{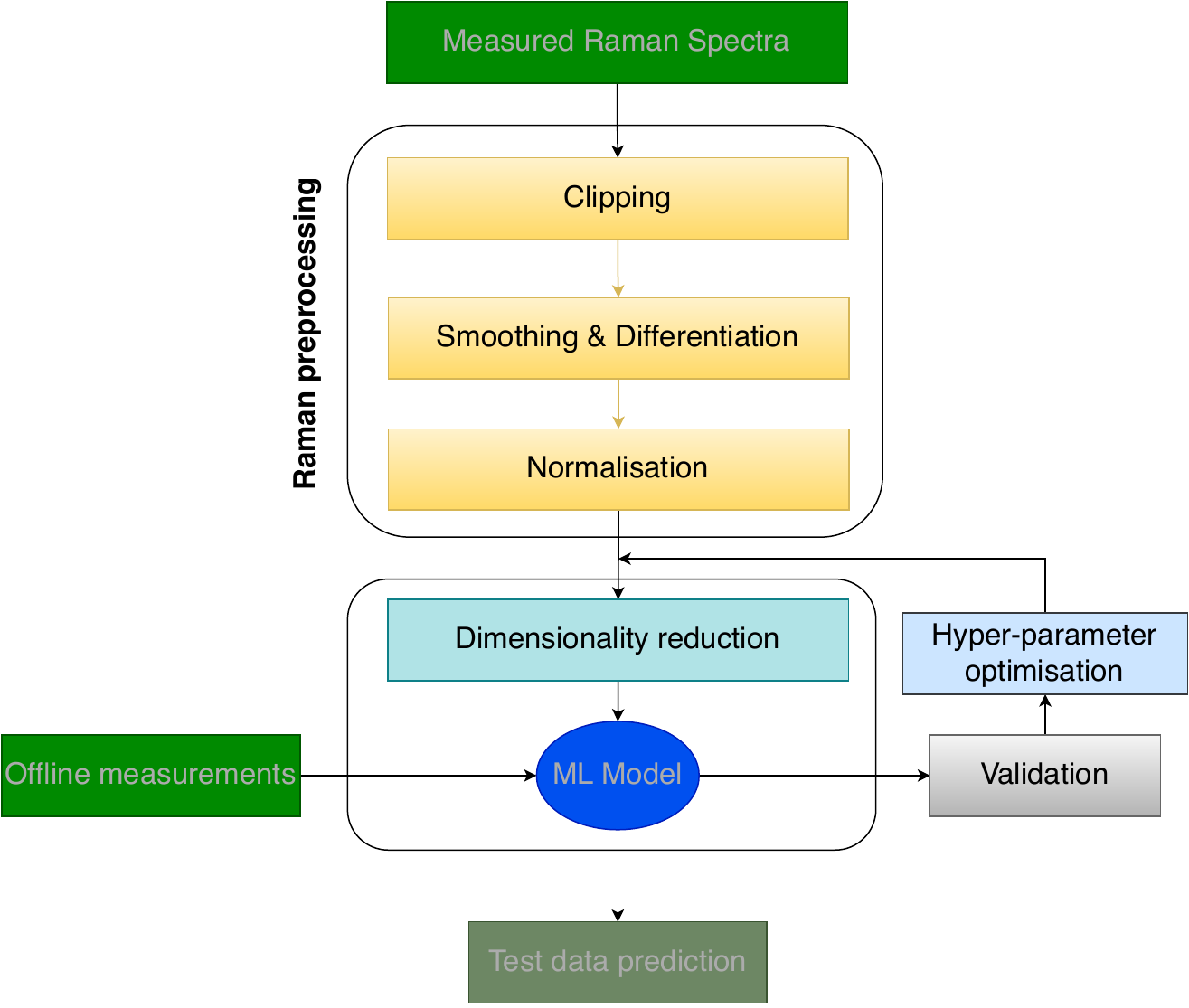}
    \caption{A pipeline for Raman spectra modeling consists of two main procedures: preprocessing and model building. The pre-processing steps aim to standardise the data by removing noise and background-related contributions. At the end of the pipeline, statistical models or machine learning approaches are constructed. These models are then assessed, and parameter optimisation may be performed based on the model outcomes. All these steps together contribute to the creation of a robust prediction from the constructed model.}
    \label{raman_pipeline}
\end{figure}

Raman spectroscopy holds great promise as a real-time monitoring tool for key analytes in mammalian cell culture fermentations. However, significant challenges accompany this promising technology, including noise, strong background fluorescence, and co-correlations between multiple components. Preprocessing of the spectra is crucial to overcome these challenges \cite{liko16} before employing multivariate regression analysis to extract relevant information and build a robust model. As affirmed by Poth et al. \cite{poma22}, preprocessing methods strongly influence the performance of machine learning models. Therefore, a typical pipeline for Raman-based machine learning models encompasses essential steps starting from Raman-pre-processing methods, as illustrated in Fig. \ref{raman_pipeline}.

Initially, the Raman spectra undergo several preprocessing steps, including wavelength selection (clipping), smoothing, signal differentiation, normalisation, and dimensionality reduction. Subsequently, the preprocessed Raman spectra, along with their corresponding offline measurements, will be employed for the development of a machine learning model. Throughout the model-building process, the hyperparameters of the models can be fine-tuned.

In this study, each Raman spectrum will be trimmed to the wavelength range of 500 to 3000 $cm^{-1}$ to eliminate highly variable spectral slopes, window peaks, an artificial jump in the Raman signal caused by spectrograph mapping on Kaiser analyzers, and interference with water \cite{poma22}. Additionally, this preprocessing step ensures the exclusion of information unrelated to glucose concentration \cite{taki23}. In this study, we employed the Savitzky-Golay procedure \cite{sago64} for the smoothing and differentiation step. This technique, based on least square fitting, was chosen for its effectiveness in preserving peaks from corruption. We used a moving average of 25 points, first-order differential, and a polynomial order of 2 to fit the samples in the Savitzky-Golay smoothing. After performing the smoothing and differentiation, the Raman spectra are standardised and, in some cases, they can be directly analysed. However, variations in intensity between Raman spectra of different samples and even within spectral maps can be significant due to changes in focusing and other experimental factors. Therefore, the use of normalisation can help alleviate this effect. Each Raman spectrum in our experiment was normalised by first subtracting the mean and then dividing by its standard deviation. 

Raman spectral datasets are typically characterised by a large number of variables, presenting challenges for statistical analysis in terms of generalisation performance and computational effort. As a result, a dimensionality reduction should be conducted before the ML model building to find a lower-dimensional representation of the original dataset without significant loss of information. Several ML models, such as PLSR, can perform this step implicitly, while many other ML models may encounter challenges when learning from a dataset with limitations in the number of samples but high dimensionality. In this study, we employed Kernel Principal Component Analysis (KPCA) as a dimensionality reduction method. The number of principal components will be fine-tuned within the range of [3, 30]. The KPCA employs the radial basis function (RBF) as a kernel, and the kernel coefficient ($\gamma$) for RBF will be a floating-point number, logarithmically tuned within the range of [$10^{-6}$, $10^2$].

\subsection{Model building}
\begin{figure}[!ht]
    \centering
    \includegraphics[width=\textwidth]{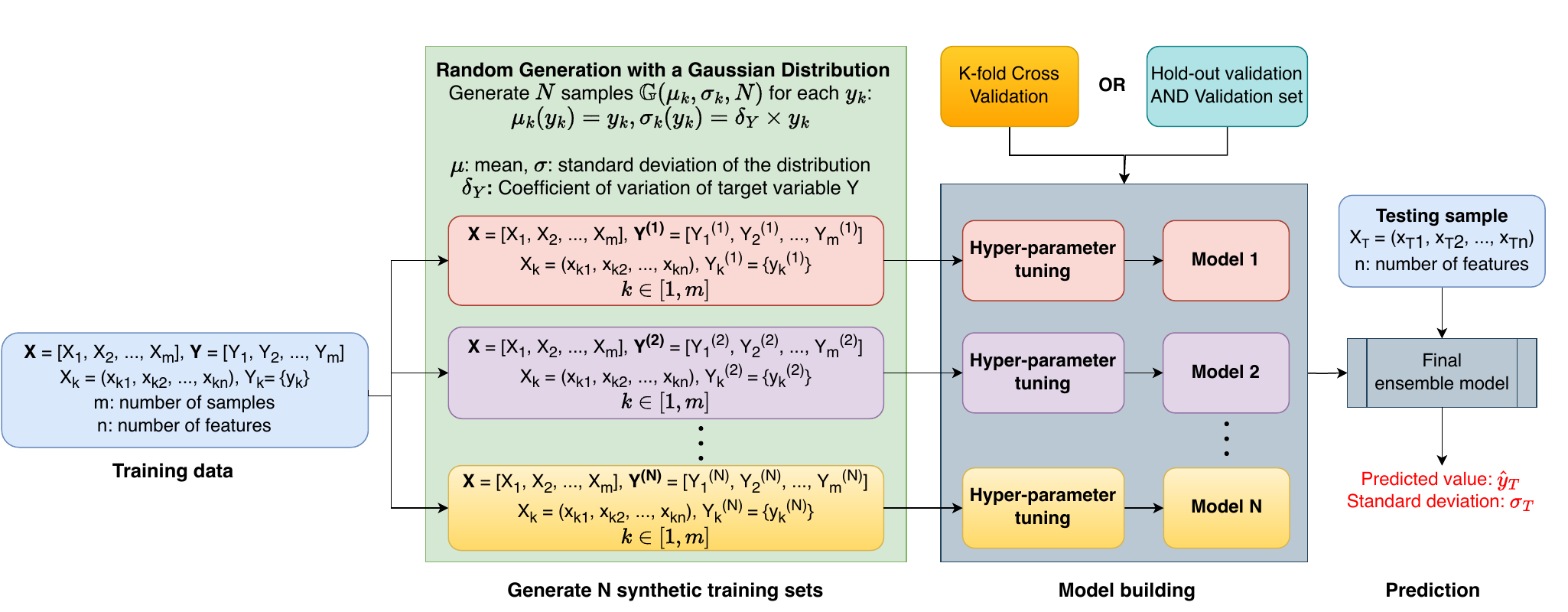}
    \caption{The modified general framework for estimating uncertainty levels of predicted values from Raman input data using ensemble learning and Monte Carlo sampling.}
    \label{framework_raman}
\end{figure}

Unlike the general framework metioned in section \ref{method}, where input features (offline measurements) are associated with coefficients of variation, the coefficients of variation of input Raman features in this problem are usually unknown because of high dimension and complex relationships among input features. We assume that only the target variable (offline glucose concentration) exhibits the uncertainty in the obtained values. Therefore, we will modify the proposed framework as in Fig. \ref{framework_raman}. In this modified framework, all base learners within the ensemble model will use the same input features but different values of the output variable. We will generate $N$ training sets $(\mathbf{X}, \mathbf{Y}^{(i)})$ ($i \in [1, N]$) for $N$ base learners by randomly generating $N$ values for each output value using Monte Carlo method with Gaussian distribution. In a mathematical form, let $\mathbf{Y} = [Y_1, Y_2,\ldots, Y_m]$ ($Y_k = \{y_k\}, y_k \in \mathbb{R}, k \in [1, m]$) be the $m$ output values in the training set. For each target value $y_k$, $N$ random values $\{ y_k^{(1)}, \ldots, y_k^{(N)}\}$ will be generated from a Gaussian distribution $\mathbb{G}(\mu_k = y_k, \sigma_k, N)$, where $\sigma_k$ is the standard deviation for each offline measurement $y_k$, calculated as follows:
\begin{equation}
\sigma_k =
\begin{cases}
\delta_Y, & \text{ if } y_k \le \beta \\
\delta_Y \times y_k, & \text{ if } y_k > \beta
\end{cases}
\end{equation}
where $\delta_Y$ is the maximum coefficient of variation of output variable $\mathbf{Y}$, $\beta$ is the threshold value used to compute the standard deviation for each offline measurement, depending on the measuring devices. For example, $\delta_Y = 0.07$ and $\beta = 1$ for glucose concentration measured by NOVA BioProfile Flex in our experiment.

After generating $N$ samples for all output values in the training set, we will concatenate all values $y_{k}^{(i)}$ at the $i^{th}$ position ($i \in [1, N]$) to create the $i^{th}$ training set $(\mathbf{X}, \mathbf{Y}^{(i)}$ in order to train the $i^{th}$ base learner within the ensemble model. As all base learners utilise the same input features, it is not advisable to set identical best hyper-parameters for all base learners, as indicated in the general framework in Fig. \ref{framework}. Instead, during the model-building process, we will conduct a hyper-parameter tuning procedure to identify the specific optimal set of hyper-parameters for each base learner, employing k-fold cross-validation or hold-out validation. In the case of using hold-out validation, a separate validation set ($\mathbf{X}^{val}, \mathbf{Y}_k^{val}$) needs to be prepared in the same manner as the training set ($\mathbf{X}, \mathbf{Y}_k$). To facilitate fine-tuning for each base learner using the Optuna library \cite{aksa19}, a fixed value of the number of base learners ($N$) needs to be used, as opposed to considering it as a tunable hyper-parameter. 

\begin{figure}[!ht]
\centering
\begin{subfloat}[5-fold CV for parameter tuning]{
\includegraphics[width=0.48\textwidth]{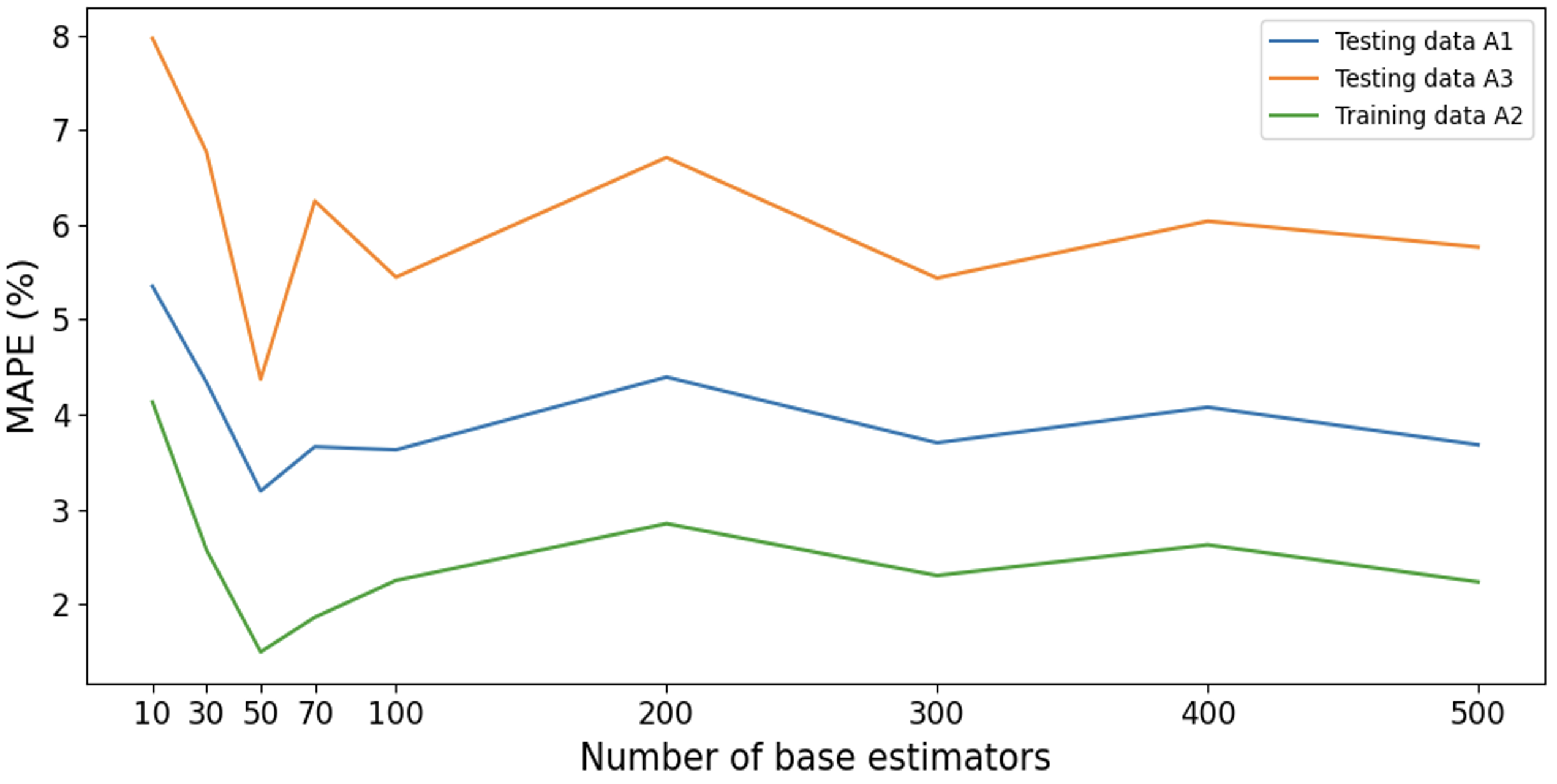}}
\end{subfloat}
\begin{subfloat}[Hold-out validation for parameter tuning]{
\includegraphics[width=0.48\textwidth]{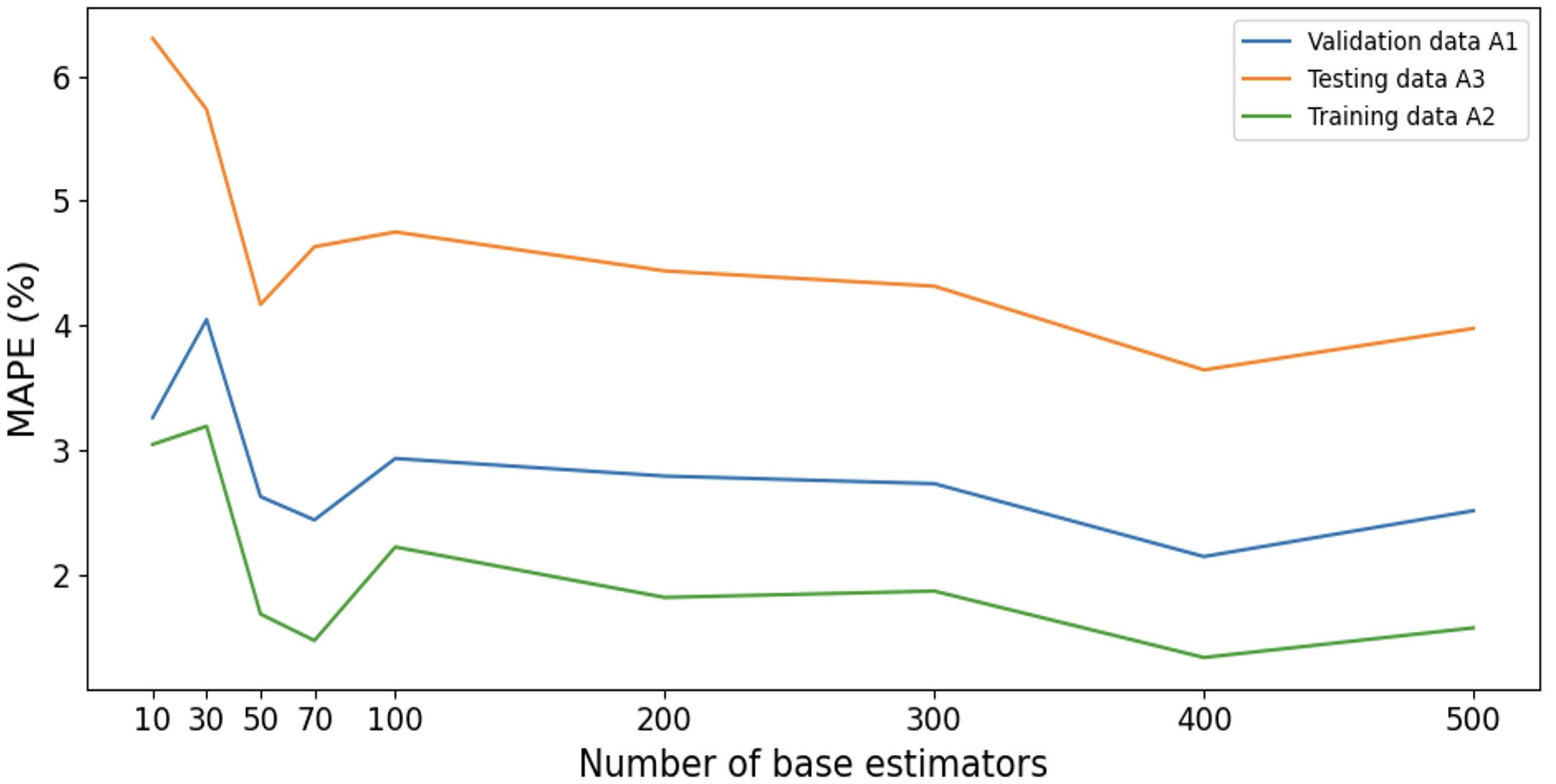}}
\end{subfloat}
\caption{The impact of the number of estimators on the prediction performance of an ensemble model consisting of KPCA and SVR base estimators.}\label{opt_N}
\end{figure}

We performed an initial experiment to identify the suitable value of $N$ using Support Vector Regression as base learners in the ensemble model and KPCA as a dimensionality reduction method. The data from bioreactor A2 was used to train the ensemble model. If the 5-fold cross-validation method is employed for hyper-parameter tuning of the base estimators, the performance of the trained model will be tested on the data from bioreactor A1 (belonging to the same project as A2) and bioreactor A3 (from a different project). If the hold-out validation approach is used for hyper-parameter tuning, the data from bioreactor A1 is used as a validation set. For this experiment, the number of base estimators considered includes 10, 30, 50, 70, 100, 200, 300, 400, and 500. The input Raman spectra were pre-processed as depicted in subsection \ref{raman_preprocessing}. The predicted performance of the trained models is presented in Fig. \ref{opt_N}.

We can observe that for small numbers of base estimators (10 and 30 estimators), the testing error is high for both the outcomes of the 5-fold CV and the hold-out validation methods. However, when the number of base estimators is equal to or greater than 50, increasing the number of base estimators does not significantly contribute to the reduction of prediction errors. Therefore, it can be concluded that 50 base estimators are sufficient to achieve good predictions without requiring a long training time in this case. As a result, we will use $N = 50$ to report the outcomes in the next parts.

For the model building step, this experiment also uses the same ML algorithms for small datasets as in the section \ref{monitoring}, including SVR, PLSR, and GP. The hyper-parameters of these algorithms will be fine-tuned using Bayesian optimisation methods within the Optuna library. Two validation methods are employed for hyper-parameter tuning: the 5-fold cross-validation and the hold-out validation. For the hold-out validation, the entire dataset of another bioreactor run would be used for validation. The experiments in this study involved 100 iterations for hyper-parameter optimisation. The ranges of hyper-parameters for each ML model are provided in Table \ref{raman_parameter}. It is noted that the single SVR model and the SVR used as a base learner within the ensemble model will use the same searching range of hyperparameter values.

\begin{table}
\centering
\renewcommand{\arraystretch}{0.1}
\begin{tabular}{|l|m{12cm}|}
\hline
\textbf{Model} & \textbf{Range} \\
\hline
SVR & \begin{itemize}[leftmargin=*, itemsep=0pt]
    \item Regularization parameter ($C$): [$10^{-5}$, $10^6$] in the logarithm domain
    \item Kernel coefficient for `rbf' kernel ($\gamma$): [$10^{-6}$, $10^2$] in the logarithm domain
    \item Epsilon parameter ($\epsilon$): [$10^{-3}$, $10^2$] in the logarithm domain
\end{itemize} \\
\hline
PLSR & \begin{itemize}[leftmargin=*, itemsep=0pt]
    \item Number of components ($n\_components$): [3, 30]
\end{itemize}  \\
\hline
GP & \begin{itemize}[leftmargin=*, itemsep=0pt]
    \item Kernel: \{DotProduct, Matern, RBF, RationalQuadratic \}
    \item The length scale or inhomogenity of the kernel: [$10^{-3}$, $10^3$] in the logarithm domain
    \item Variance of additional Gaussian measurement noise ($\alpha$): [$10^{-8}$, $10^{-1}$] in the logarithm domain
\end{itemize}  \\
\hline
\end{tabular}
\caption{The search range of hyper-parameters for Raman-based ML models.} \label{raman_parameter}
\end{table}

\subsection{Evaluation metrics}
In this experiment, we would like to assess the average percentage difference between predicted and actual values to compare with the maximum coefficient of variation of the actual values (about 7\% for glucose concentrations). Therefore, the metric employed to assess the performance of ML models and determine the optimal parameter configurations in this section is the Mean Absolute Percentage Error (MAPE). Additionally, MAPE is scale-independent, meaning it provides a percentage error and is not affected by the scale of the data. This makes it useful when comparing the performance of different ML models on testing data of bioreactors A1 and A3 with different scales. The MAPE metric will be computed as follows:
\begin{equation}
    MAPE = \cfrac{1}{N_{test}} \cdot \sum_{i=1}^{N_{test}}{\left|\cfrac{\hat{y}_i - y_i}{y_i}\right|}
\end{equation}
where $N_{test}$ is the number of testing samples, $\hat{y}_i$ is the prediction of the $i^{th}$ testing sample, and $y_i$ is the true value of the $i^{th}$ testing sample. For the ML models which are able to provide the standard deviation associated with the predictive values, we will compute both MAPE scores for the upper bound ($\hat{y} + 2\sigma$) and the lower bound ($\hat{y} - 2\sigma$) as follows:
\begin{equation}
    MAPE^+ = \cfrac{1}{N_{test}} \cdot \sum_{i=1}^{N_{test}}{\left|\cfrac{(\hat{y}_i + 2\sigma_i) - y_i}{y_i}\right|}
\end{equation}
\begin{equation}
    MAPE^- = \cfrac{1}{N_{test}} \cdot \sum_{i=1}^{N_{test}}{\left|\cfrac{(\hat{y}_i - 2\sigma_i) - y_i}{y_i}\right|}
\end{equation}
where $\sigma_i$ is the standard deviation associated with the prediction $\hat{y}_i$ of the $i^{th}$ testing sample. If the $MAPE$ score is small, while the values of $MAPE^+$ and $MAPE^-$ are high, we can confirm that the average of predictions provided by all individual base learners contributes to the mitigation of variations among individual learners and increasing in the accuracy.

\subsection{Analytical results}
This section will present the practical outcomes of glucose concentration prediction employing various ML models. These models encompass the combination of KPCA and SVR, the combination of KPCA and GP, PLSR, ensemble of KPCA and SVRs, and ensemble of PLSRs. The training dataset is derived from bioreactor A2, while datasets from bioreactors A1 and A3 serve as the testing data. In the case of using hold-out validation, the dataset from bioreactor A1 is utilized as the validation set.

\subsubsection{Comparing the uncertainty level in the predicted results with the same and different projects}
This section aims to evaluate the uncertainty level of real-time predictions made by the ensemble of KPCA and SVR base estimators. The predictions are based on Raman spectra that have undergone preprocessing steps as presented in section \ref{raman_preprocessing}.

\begin{figure}[!ht]
\centering
\begin{subfloat}[5-fold CV - Bioreactor A1 (same project)]{
\includegraphics[width=0.48\textwidth]{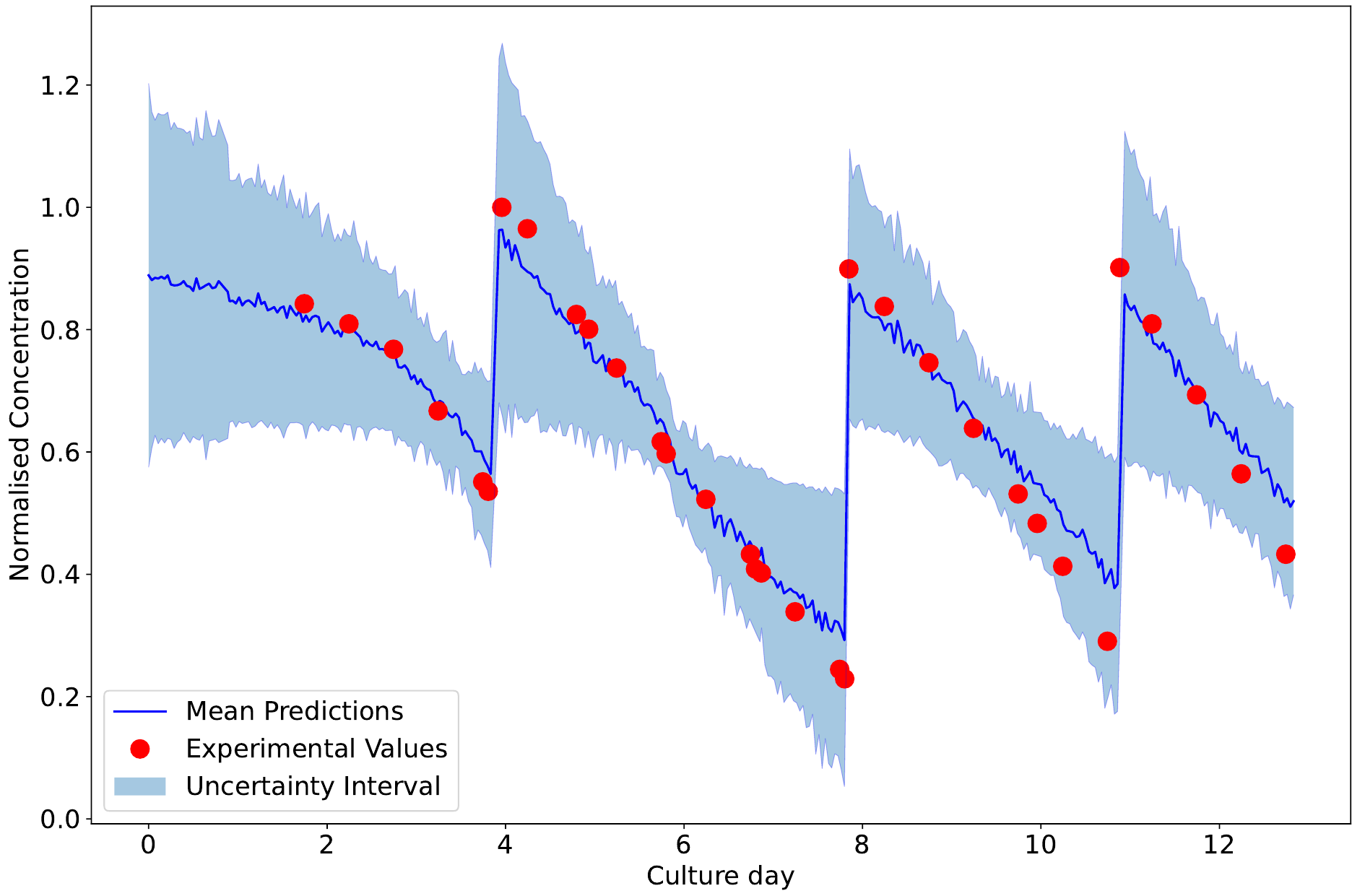}}
\end{subfloat}
\begin{subfloat}[5-fold CV - Bioreactor A3 (different project)]{
\includegraphics[width=0.48\textwidth]{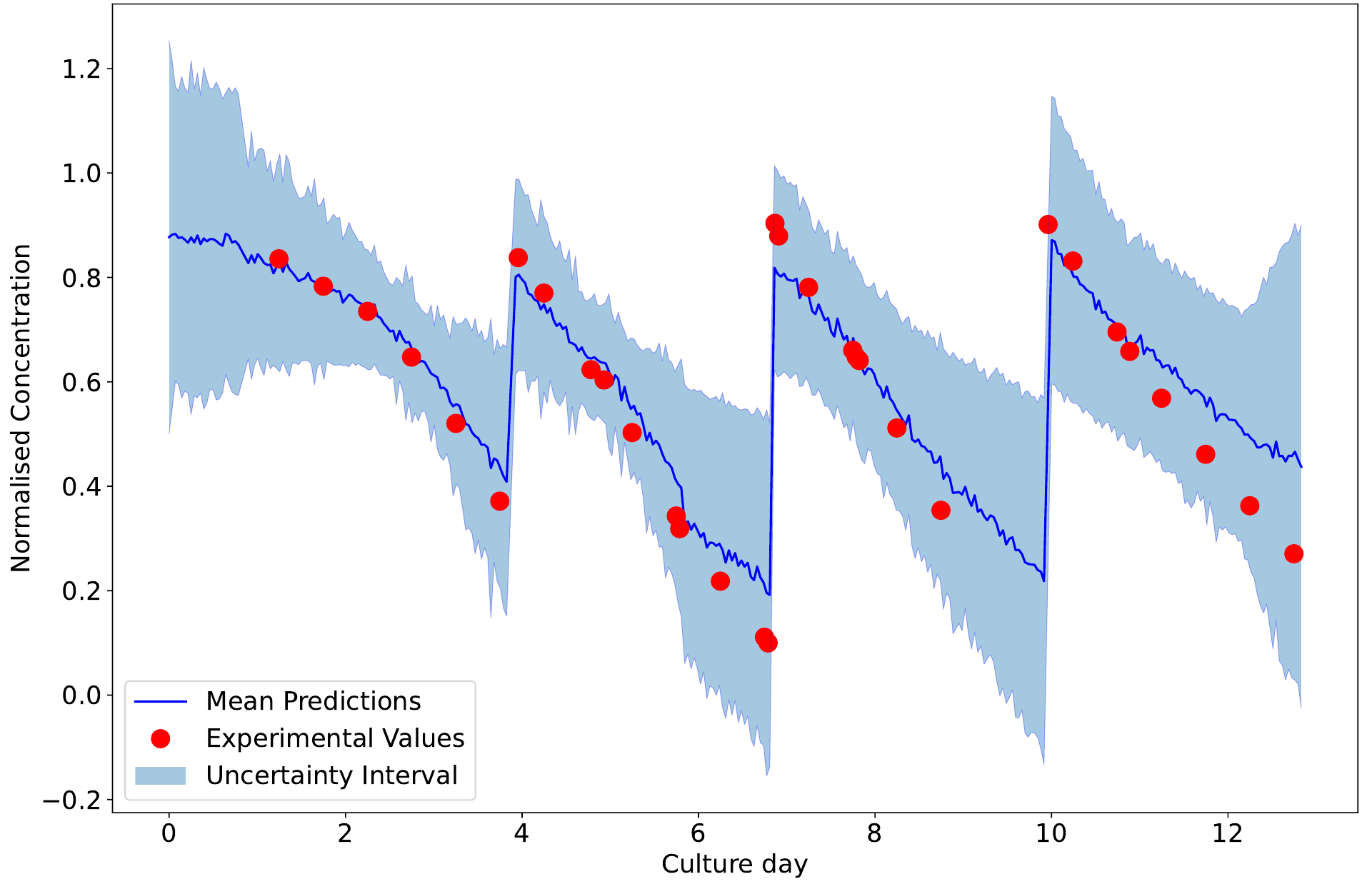}}
\end{subfloat}
\begin{subfloat}[Hold-out validation - Bioreactor A1 (validation set)]{
\includegraphics[width=0.48\textwidth]{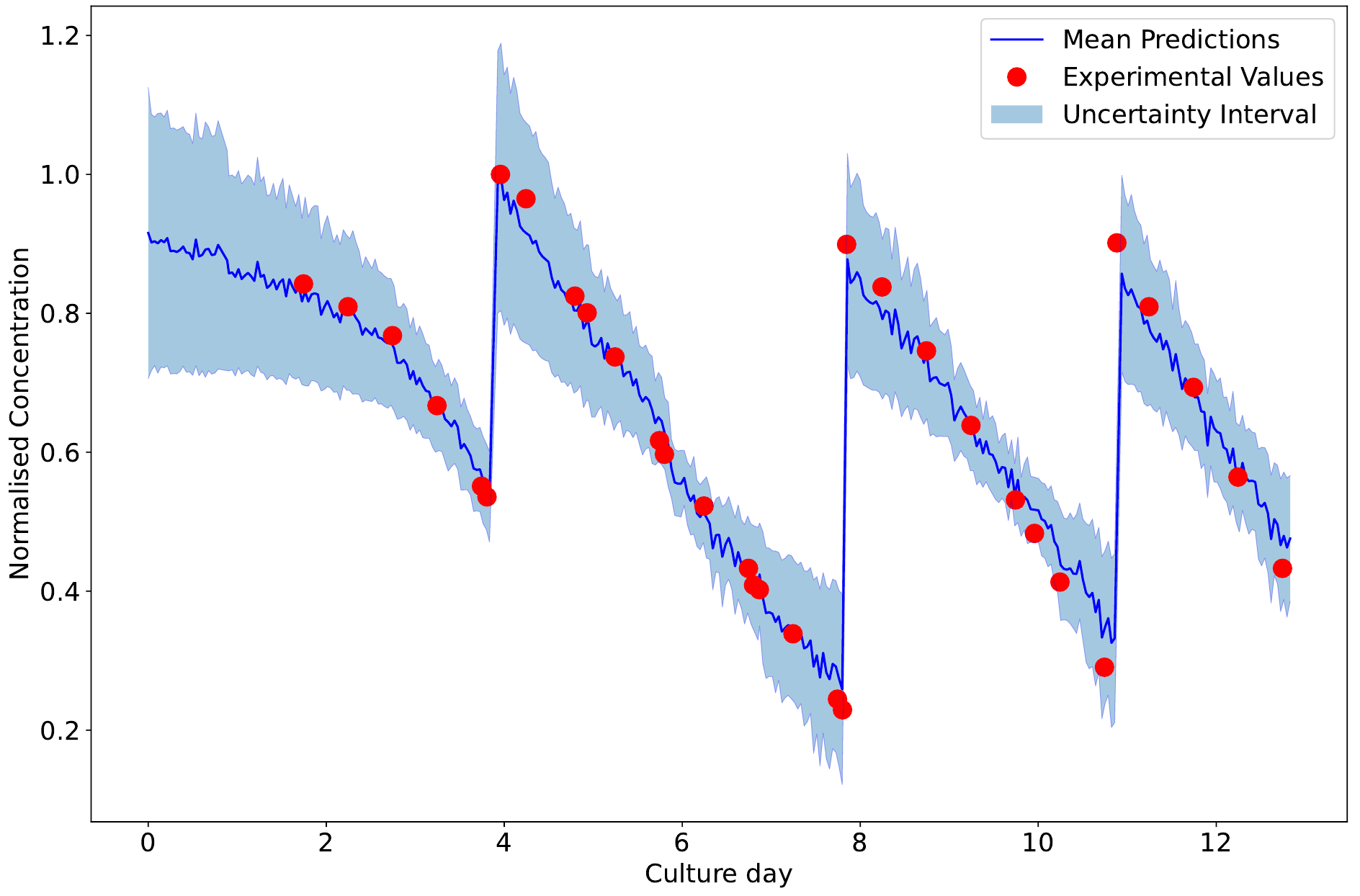}}
\end{subfloat}
\begin{subfloat}[Hold-out validation - Bioreactor A3]{
\includegraphics[width=0.48\textwidth]{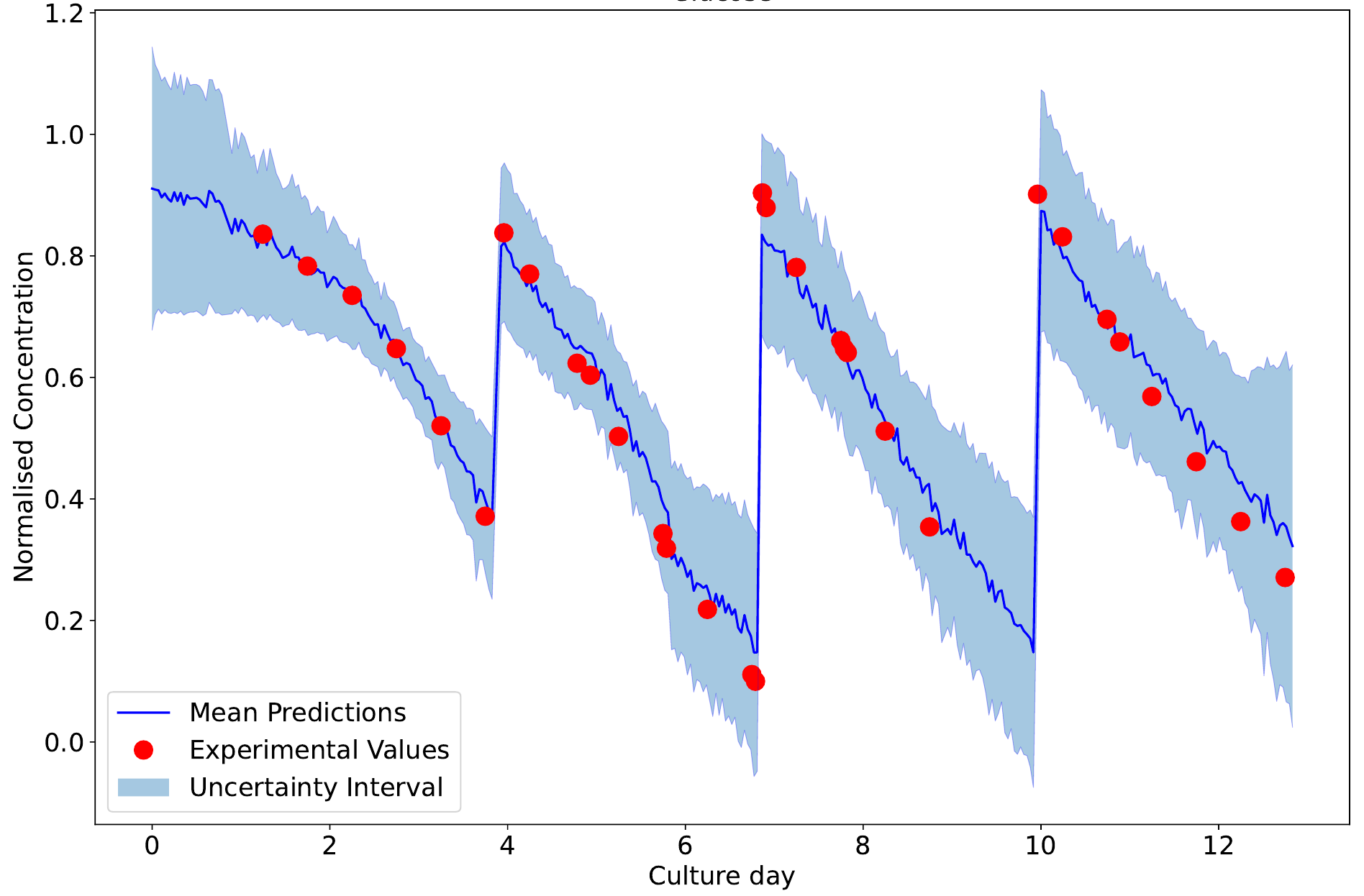}}
\end{subfloat}
\caption{Real-time predictions of glucose concentrations within various bioreactors with the same and different projects using an ensemble of KPCA and SVRs.} \label{realtime_uncertainty}
\end{figure}

Fig. \ref{realtime_uncertainty} illustrates the real-time predictions and uncertainty levels of glucose concentrations for different bioreactors within the same and different projects using an ensemble model. This is a typical use-case in industry for a new project which will initially have very little specific data available for training, relying instead on data from previous projects. When considering both validation methods, we observe that the uncertainty levels of predicted values within the bioreactors with the same project are smaller compared to those of different projects. This discrepancy arises from the differences in metabolism and growth between cell lines which is reflected in the Raman spectra. These elements are quite similar between cell culture bioreactors within the same project, so the information recorded in the Raman spectra of the training data is more likely to be present in the testing data of the same project. In contrast, cell lines from different projects can have distinct interactions with the culture composition and so the information recorded in the Raman spectra testing data may not be available if the training data is from a different project. Consequently, this disparity impacts the predicted values for the testing data and increases the uncertainty levels. To mitigate this, we can reduce the uncertainty in predicted values for testing data from a different project by validating and optimising the parameters of the trained model on the validation data within the same project as testing data. 

Furthermore, as illustrated in Fig. \ref{realtime_uncertainty}, the uncertainty associated with real-time predictions of the ensemble models fine-tuned using 5-fold cross-validation is significantly higher than that of the ensemble model fine-tuned using hold-out validation. This observation underscores that having a validation set encompassing data from all cell culture days enhances the reliability of the ensemble model, whereas validation on only a limited portion of cell culture days tends to amplify the uncertainty in predictive outcomes. 

\subsubsection{Comparing the predicted performance among the different, tested ML models}
This section aims to compare the predicted performance of the proposed method with two different sets of base estimators: PLSR and the combination of KPCA and SVR, to the performance of single models such as PLSR, the combination of KPCA and GP, and the combination of KPCA and SVR. Since the base estimators of the proposed ensemble method were trained on data with the target variable within a 7\% deviation from the ground truth values, we do not expect the ensemble method to produce the best predicted performance compared to those trained on the ground truth values. However, we anticipate that the performance of the ensemble model will be comparable to that of the best single model. One of the strengths of the proposed method, compared to the single ML models, is its ability to provide the standard deviation of the predicted outcomes based on the coefficient of variation of bio-analysers.

In this experiment, we used 50 base estimators to build the ensemble model. The ML models were trained using the data from the bioreactor A2. For the case of 5-fold CV employed for the hyper-parameter tuning, the trained models were tested on the data from the bioreactor A1 (the same project as A2) and the bioreactor A3 (a different project from A2). In the case of using hold-out validation for the hyper-parameter tuning, the data from bioreactor A1 was used as a validation set, and the data from the bioreactor A3 was used as a testing set. 

\begin{table}[!ht]
\centering
\begin{tabular}{|m{4cm}|c|c|c|c|c|c|}
\hline
\multirow{2}{*}{\textbf{MAPE(\%)}} & \multicolumn{3}{c|}{\textbf{Bioreactor A1}} &  \multicolumn{3}{c|}{\textbf{Bioreactor A3}}   \\
\cline{2-7}
 & Prediction & Upper bound & Lower bound & Prediction & Upper bound & Lower bound \\
\hline
Ensemble of KPCA + SVRs & 3.8280 & 18.1349 & 13.8381 & 6.1250 & 28.4885 & 19.5367 \\
\hline
Ensemble of PLSRs & 4.5388 & 14.0537 & 6.2497 & 4.3285 & 13.1566 & 11.2454 \\
\hline
KPCA + SVR & 1.6133	& - & - & 2.5731 & - & - \\
\hline
PLSR & 4.2982 & - & - & 4.2428 & - & - \\
\hline
KPCA + GP & 2.0874 & 2.0942 & 2.0806 & 4.1223 & 4.1400 & 4.1047 \\
\hline
\end{tabular}
\caption{The MAPE (\%) values of various ML models trained and optimised by the 5-fold CV method.} \label{result_glucose_cv}
\end{table}

Table \ref{result_glucose_cv} presents the predicted performance of various ML models on the data from the bioreactor A1 (the same project as A2) and the bioreactor A3 (a different project from A2) using the 5-fold CV approach. Among the models, the combination of KPCA and SVR provided the best performance on predictions of glucose concentrations of the bioreactor A3, while the combination of KPCA and GP yielded the best results on the bioreactor A1 of the same project with the training data.

In this experiment, the ensemble of models did not consistently outperform the single models in terms of predicted values. While the ensemble of PLSRs can outperform a single PLSR model, the ensemble of KPCA and SVRs is not able to provide a better performance in comparison of the single combination of KPCA and SVR. However, the difference in predicted performance between the ensemble models and single models was generally within an error of around 4\%. Nevertheless, the ensemble model provided additional benefits by generating an estimation of the uncertainty for each predicted outcome compared to the single models. It is worth noting that the prediction errors of the ML models in this experiment were often below 7\%, which is also smaller than the maximum error of 7\% associated with the bioanalyser for offline glucose concentrations. This observation suggests that the ML models can be effectively employed for developing soft sensors for real-time monitoring of glucose concentrations.

 \begin{table}[!ht]
\centering
\begin{tabular}{|m{4cm}|c|c|c|c|c|c|}
\hline
\multirow{2}{*}{\textbf{MAPE(\%)}} & \multicolumn{3}{c|}{\textbf{Bioreactor A1} (Validation set)} &  \multicolumn{3}{c|}{\textbf{Bioreactor A3}}   \\
\cline{2-7}
 & Prediction & Upper bound & Lower bound & Prediction & Upper bound & Lower bound \\
\hline
Ensemble of KPCA + SVRs & 2.1227 & 9.8612 & 8.9721 & 3.6011 & 18.1493 & 14.0127 \\
\hline
Ensemble of PLSRs & 3.8046 & 13.6509	& 7.3930 & 4.2195 & 13.4981 & 12.3385 \\
\hline
KPCA + SVR & 1.3246	& - & - & 2.3634 & - & - \\
\hline
PLSR & 4.2982 & - & - & 4.2428 & - & - \\
\hline
KPCA + GP & 1.8168 & 1.8142 & 1.8194 & 3.4109 & 3.4381 & 3.3837 \\
\hline
\end{tabular}
\caption{The MAPE (\%) values of various ML models trained and optimised by the hold-out validation method.} \label{result_glucose_holdout}
\end{table}

Table \ref{result_glucose_holdout} presents the predicted performance of various ML models trained on data from the bioreactor A2, validated on data from the bioreactor A1, and tested on data from the bioreactor A3. In this scenario, the combination of KPCA and SVR stands out as the top-performing model on the testing data (A3). Furthermore, it is noticeable that the testing errors of all five ML models fine-tuned by the hold-out validation are smaller than those fine-tuned by the 5-fold cross-validation method. Additionally, the uncertainty levels on the testing data for the ensemble models using the hold-out validation are smaller than those using the 5-fold cross-validation method.

\begin{figure}[!ht]
\centering
\begin{subfloat}[KPCA + SVR - Bioreactor A1]{
\includegraphics[width=0.48\textwidth]{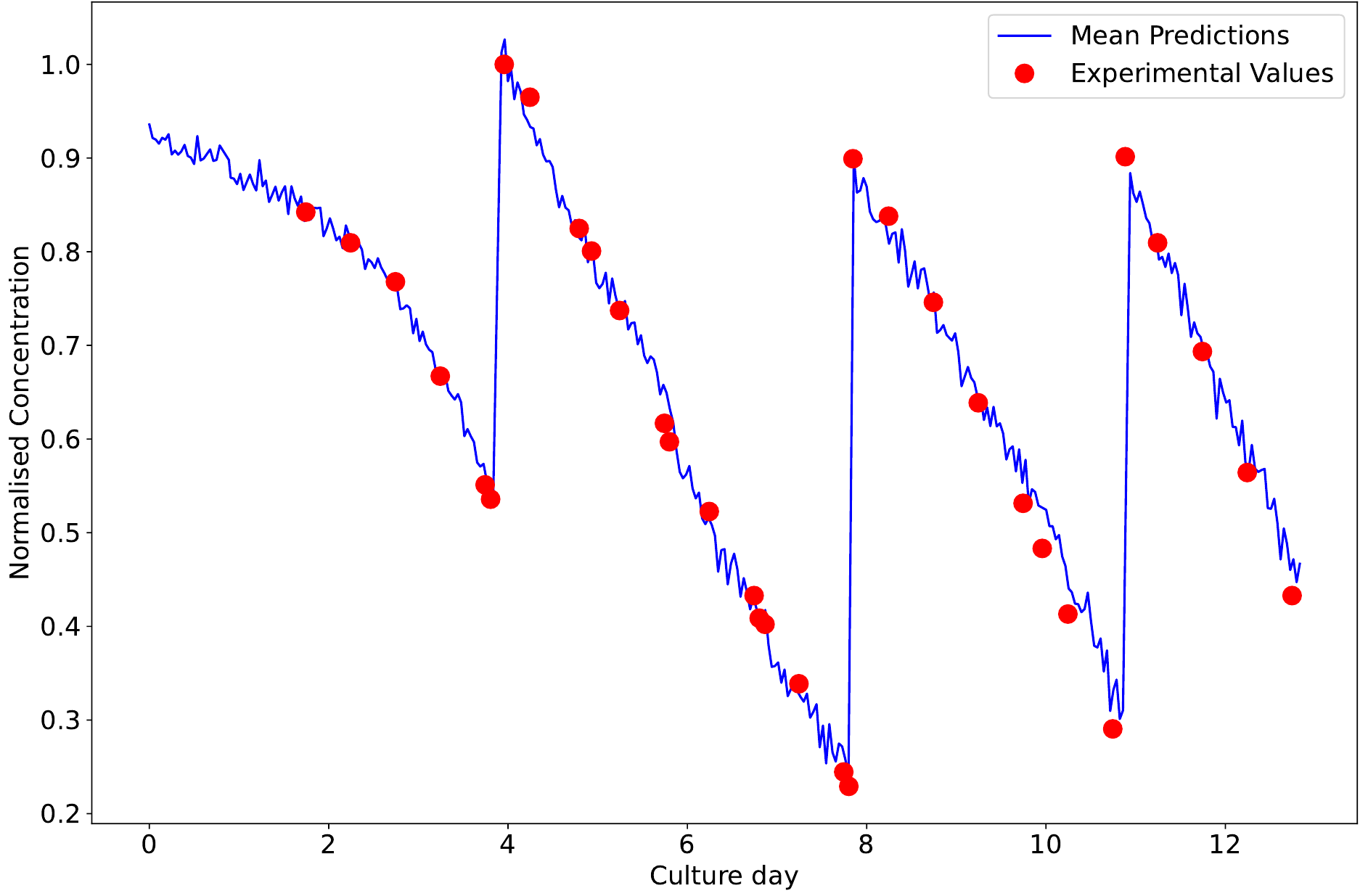}}
\end{subfloat}
\begin{subfloat}[KPCA + SVR - Bioreactor A3]{
\includegraphics[width=0.48\textwidth]{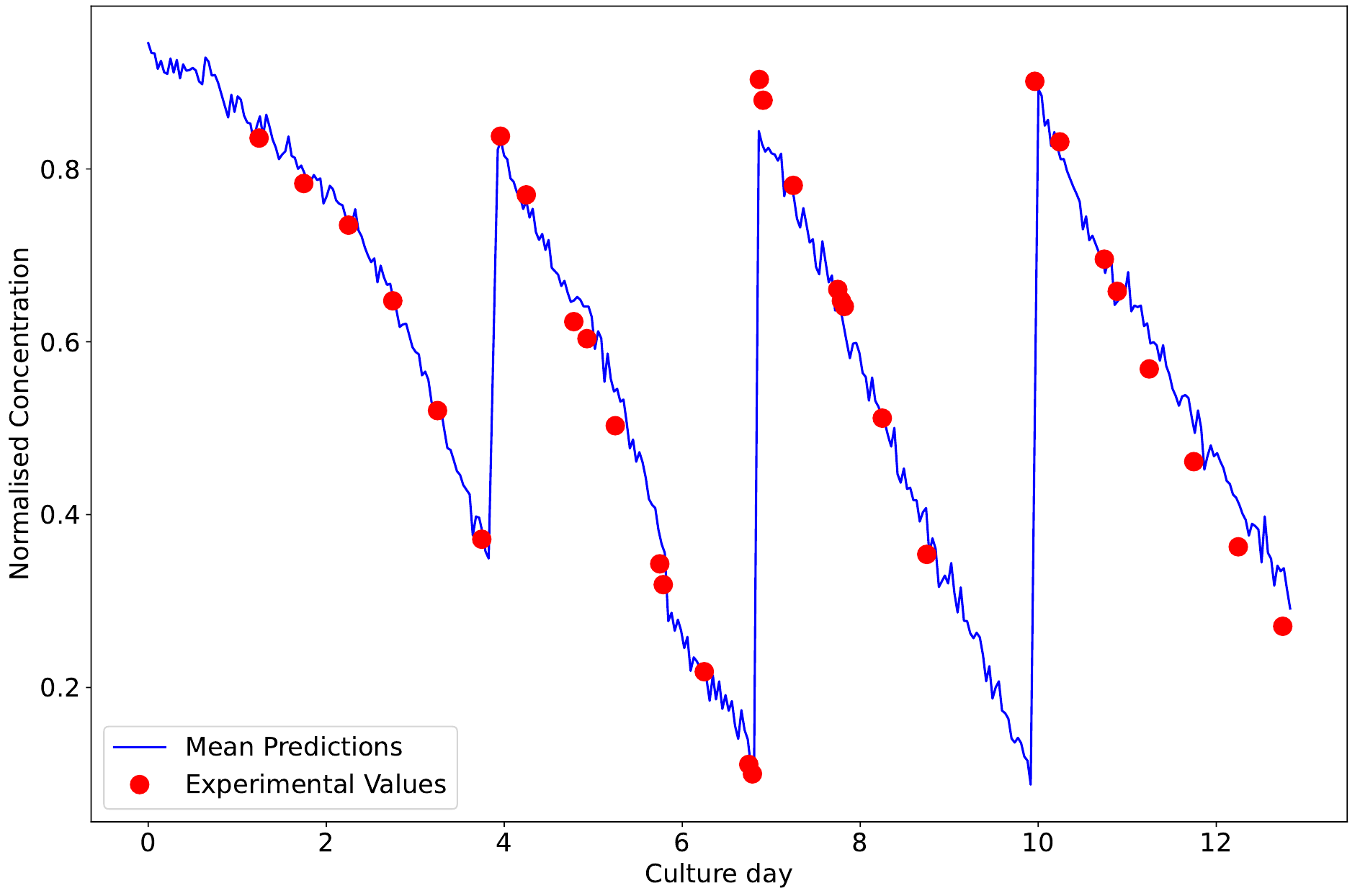}}
\end{subfloat}
\begin{subfloat}[Ensemble of KPCA and SVRs - Bioreactor A1]{
\includegraphics[width=0.48\textwidth]{glucose_ensemble_kpca_svr_cv_a1.pdf}}
\end{subfloat}
\begin{subfloat}[Ensemble of KPCA and SVRs - Bioreactor A3]{
\includegraphics[width=0.48\textwidth]{glucose_ensemble_kpca_svr_cv_a3.pdf}}
\end{subfloat}
\begin{subfloat}[PLSR - Bioreactor A1]{
\includegraphics[width=0.48\textwidth]{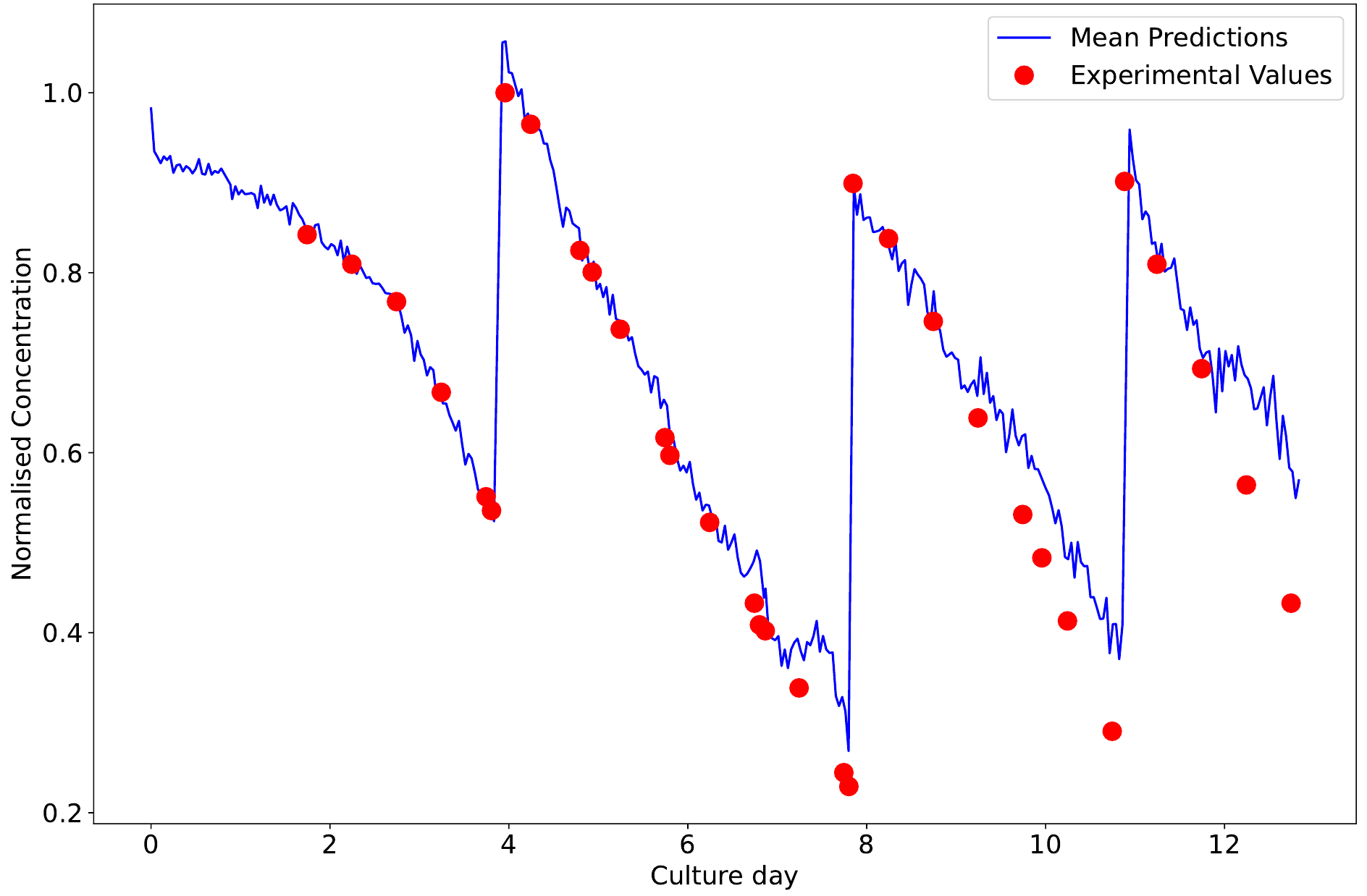}}
\end{subfloat}
\begin{subfloat}[PLSR - Bioreactor A3]{
\includegraphics[width=0.48\textwidth]{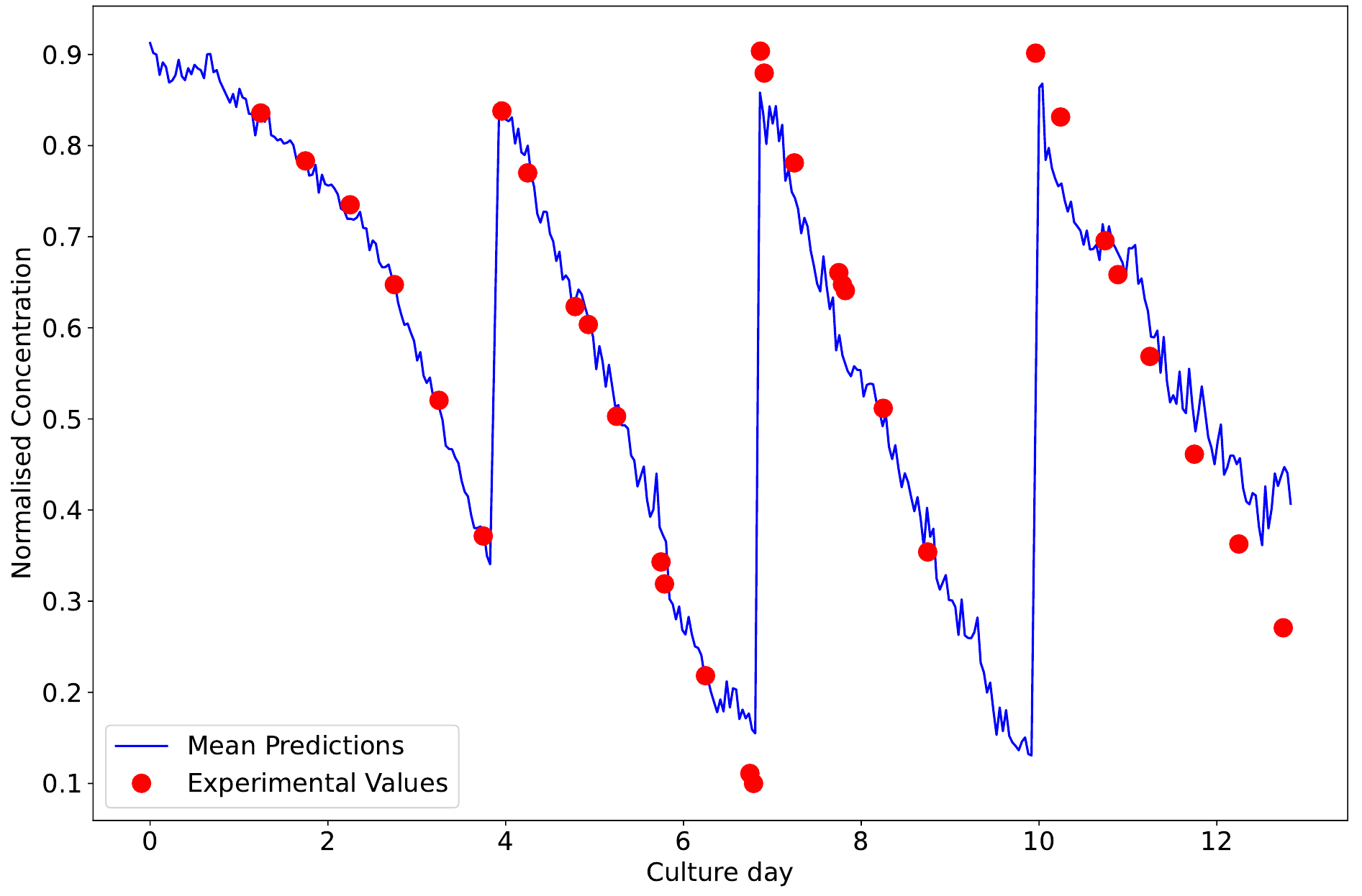}}
\end{subfloat}
\caption{Real-time monitoring of glucose concentrations in bioreactors A1 and A3 using various ML models trained on the data from bioreactor A2, with 5-fold CV used for hyper-parameter tuning.} 
\end{figure}
\pagebreak
\begin{figure}[!ht]\ContinuedFloat
\begin{subfloat}[Ensemble of PLSRs - Bioreactor A1]{
\includegraphics[width=0.48\textwidth]{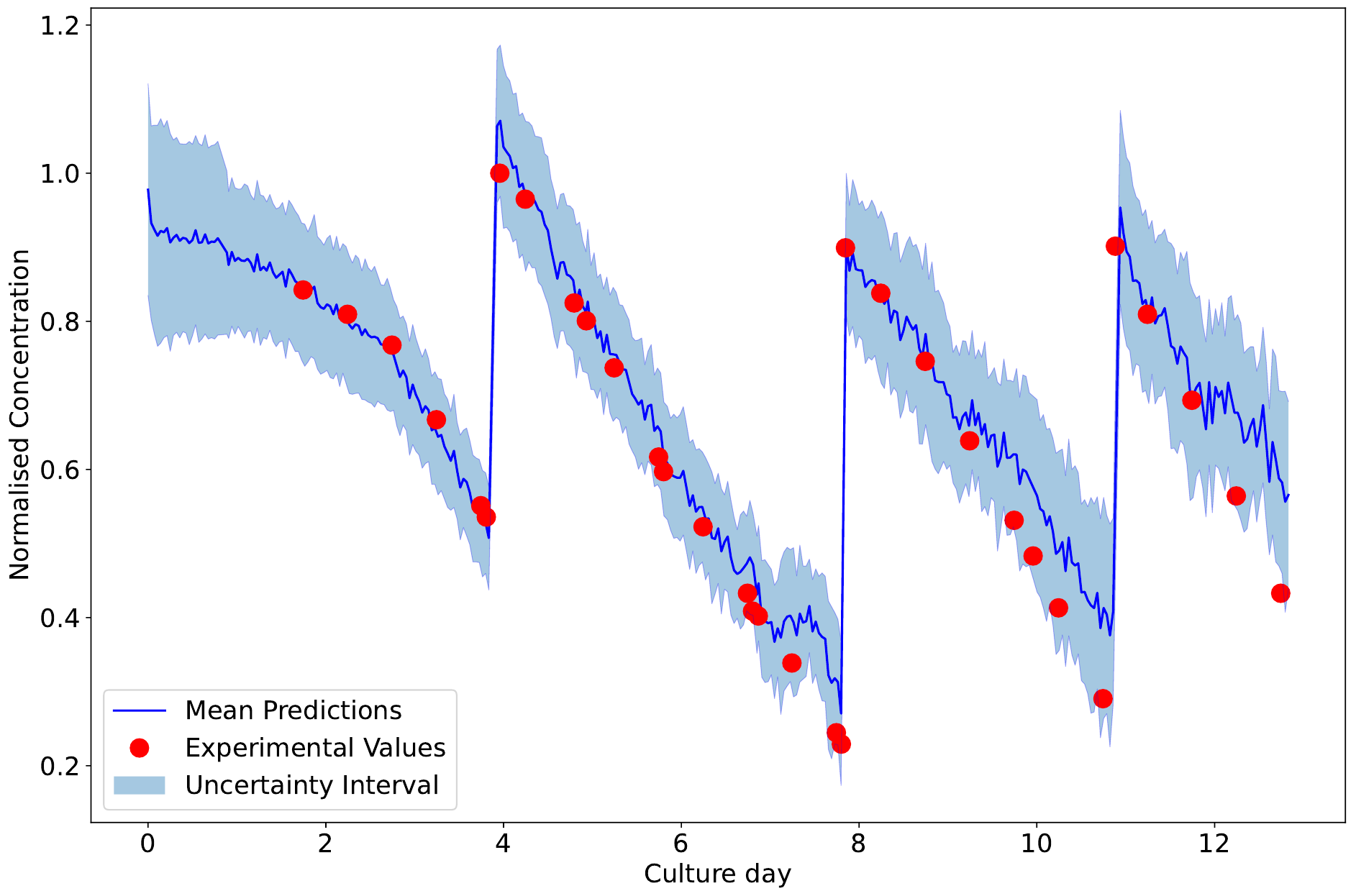}}
\end{subfloat}
\begin{subfloat}[Ensemble of PLSRs - Bioreactor A3]{
\includegraphics[width=0.48\textwidth]{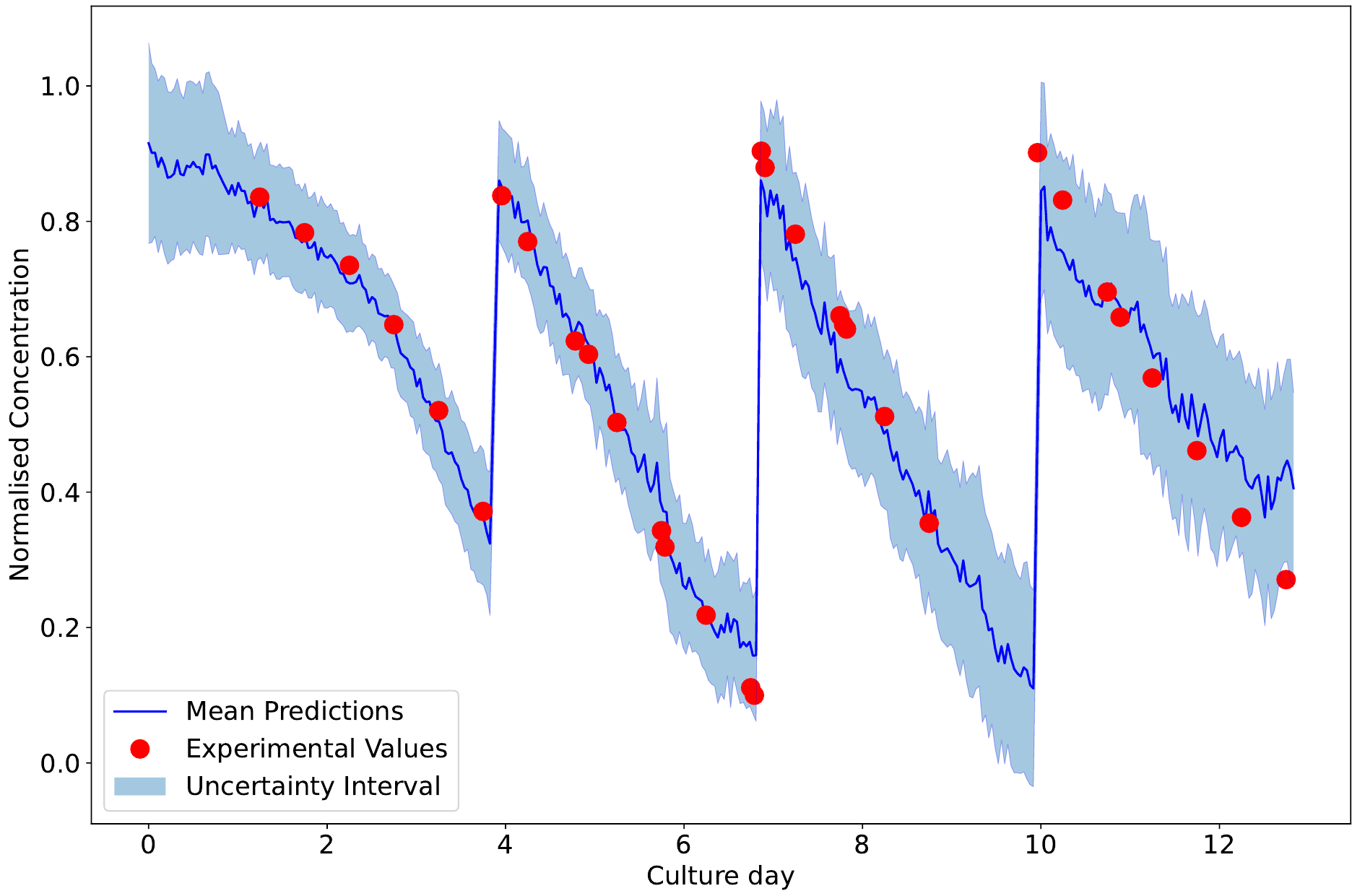}}
\end{subfloat}
\begin{subfloat}[KPCA + GP - Bioreactor A1]{
\includegraphics[width=0.48\textwidth]{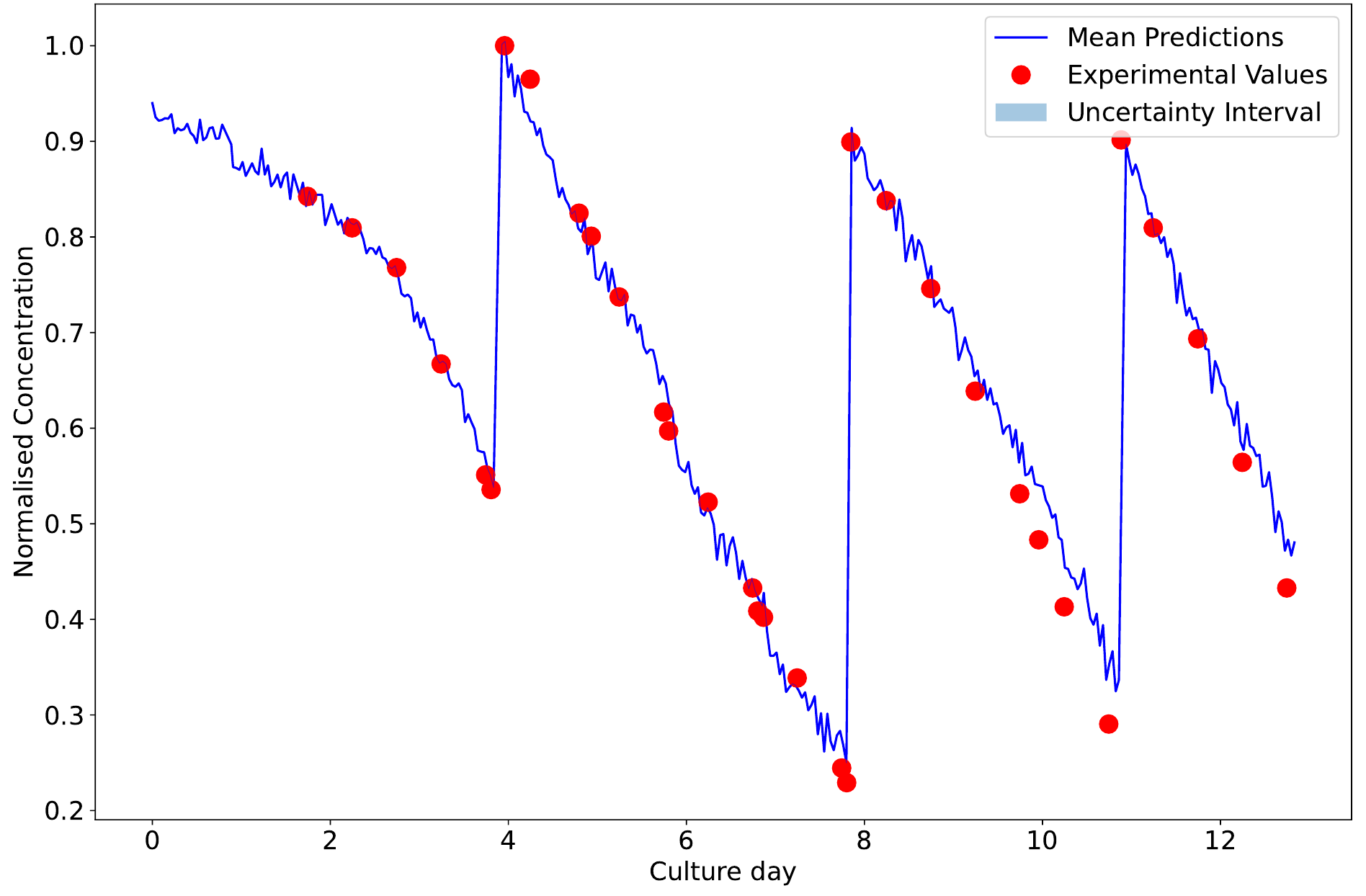}}
\end{subfloat}
\begin{subfloat}[KPCA + GP - Bioreactor A3]{
\includegraphics[width=0.48\textwidth]{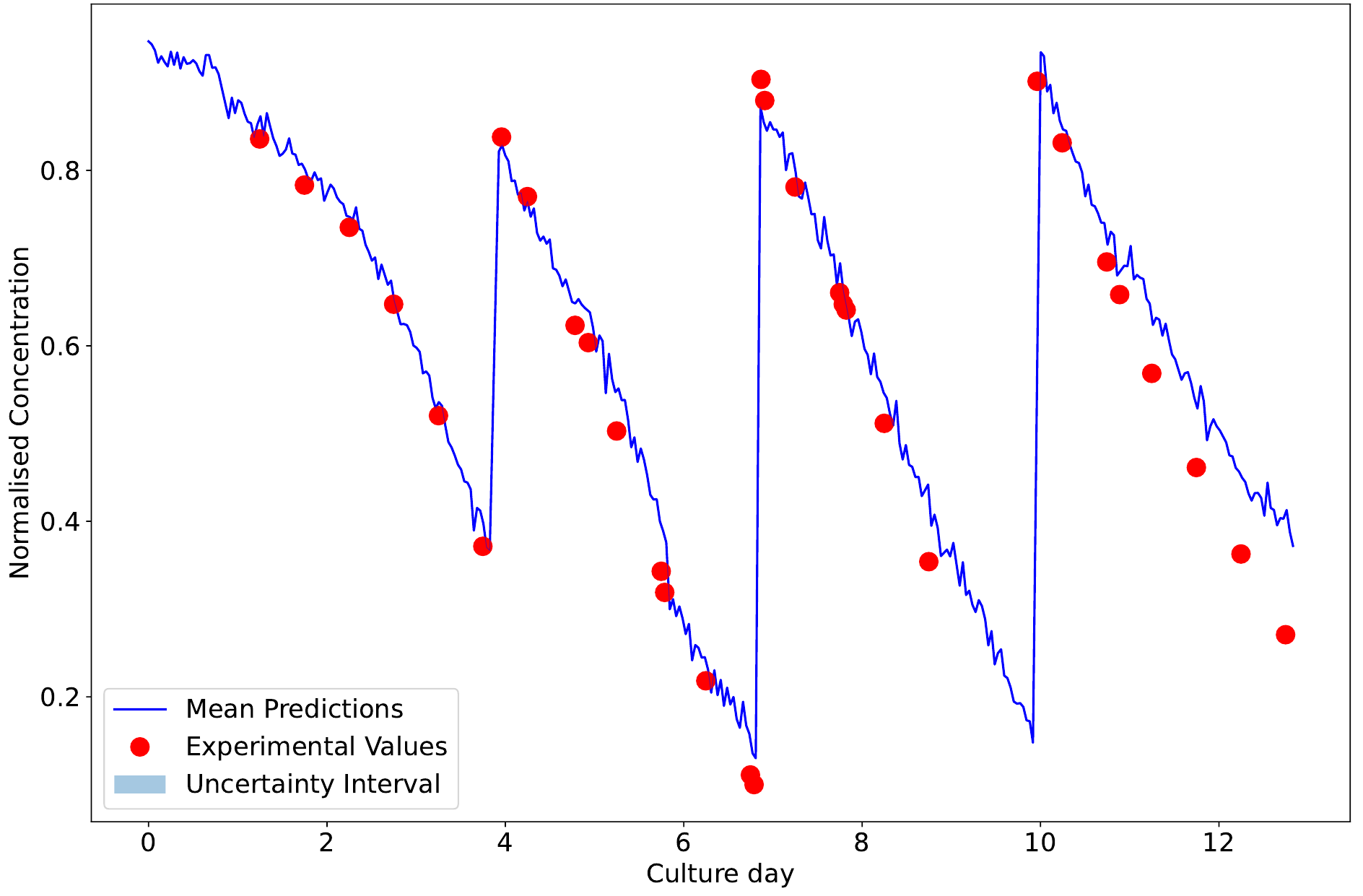}}
\end{subfloat}
\caption{Real-time monitoring of glucose concentrations in bioreactors A1 and A3 using various ML models trained on the data from bioreactor A2, with 5-fold CV used for hyper-parameter tuning (Continued).} \label{realtime_cv}
\end{figure}

\begin{figure}[!ht]
\centering
\begin{subfloat}[KPCA + SVR - Bioreactor A3]{
\includegraphics[width=0.48\textwidth]{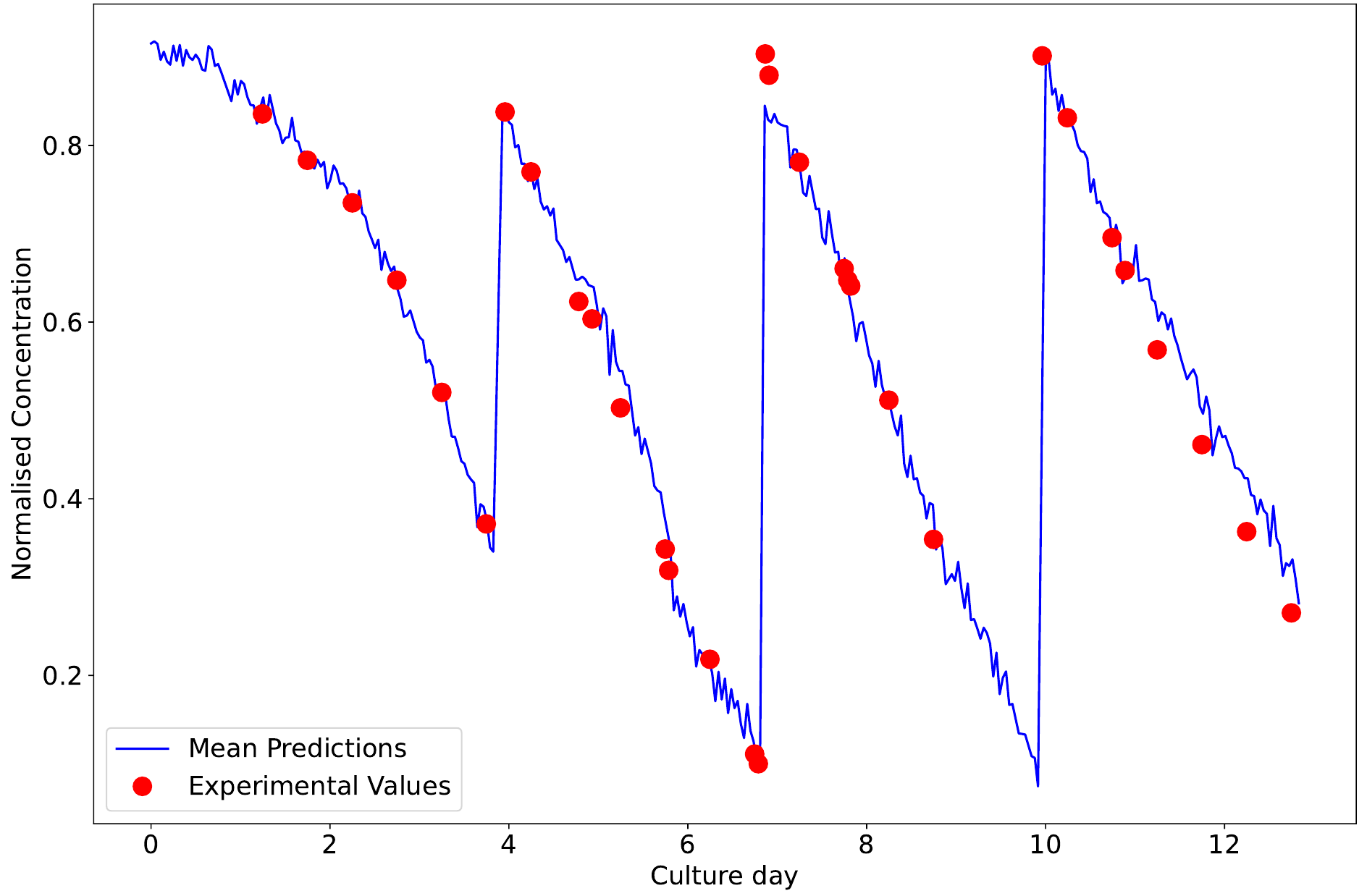}}
\end{subfloat}
\begin{subfloat}[PLSR - Bioreactor A3]{
\includegraphics[width=0.48\textwidth]{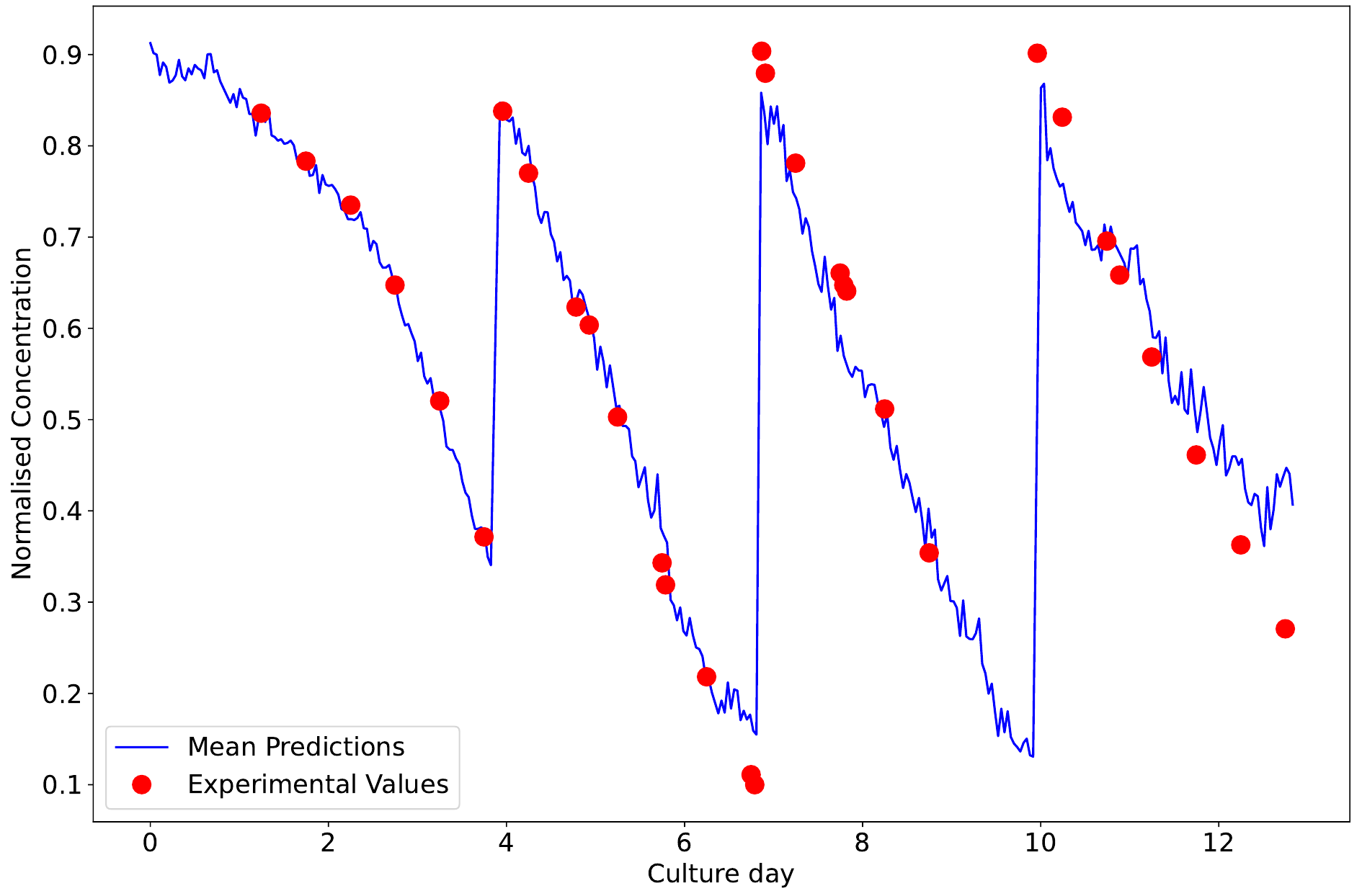}}
\end{subfloat}
\begin{subfloat}[Ensemble of KPCA and SVRs - Bioreactor A3]{
\includegraphics[width=0.48\textwidth]{glucose_ensemble_kpca_svr_holdout_a3.pdf}}
\end{subfloat}
\begin{subfloat}[Ensemble of PLSRs - Bioreactor A3]{
\includegraphics[width=0.48\textwidth]{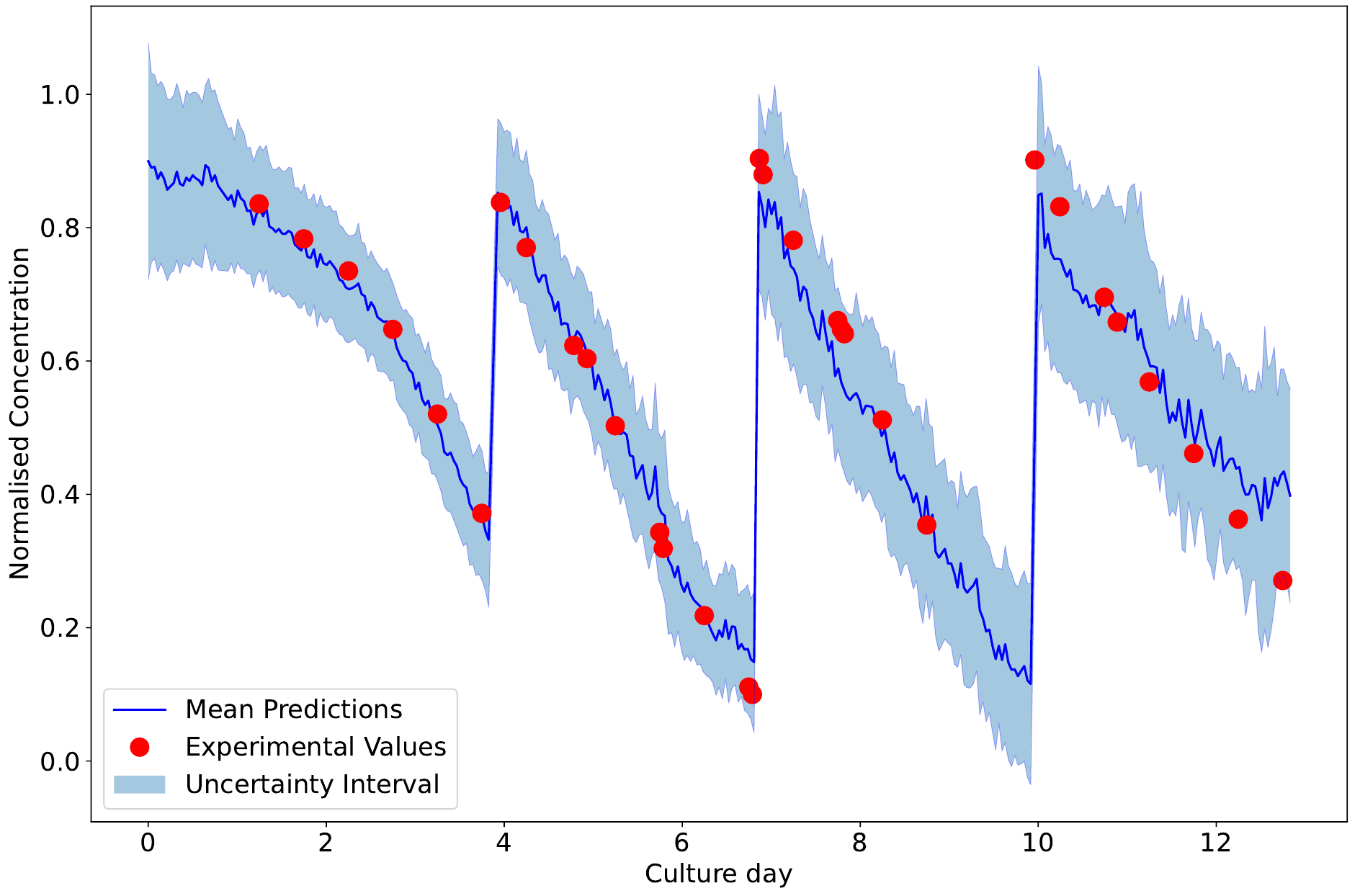}}
\end{subfloat}
\begin{subfloat}[KPCA + GP - Bioreactor A3]{
\includegraphics[width=0.48\textwidth]{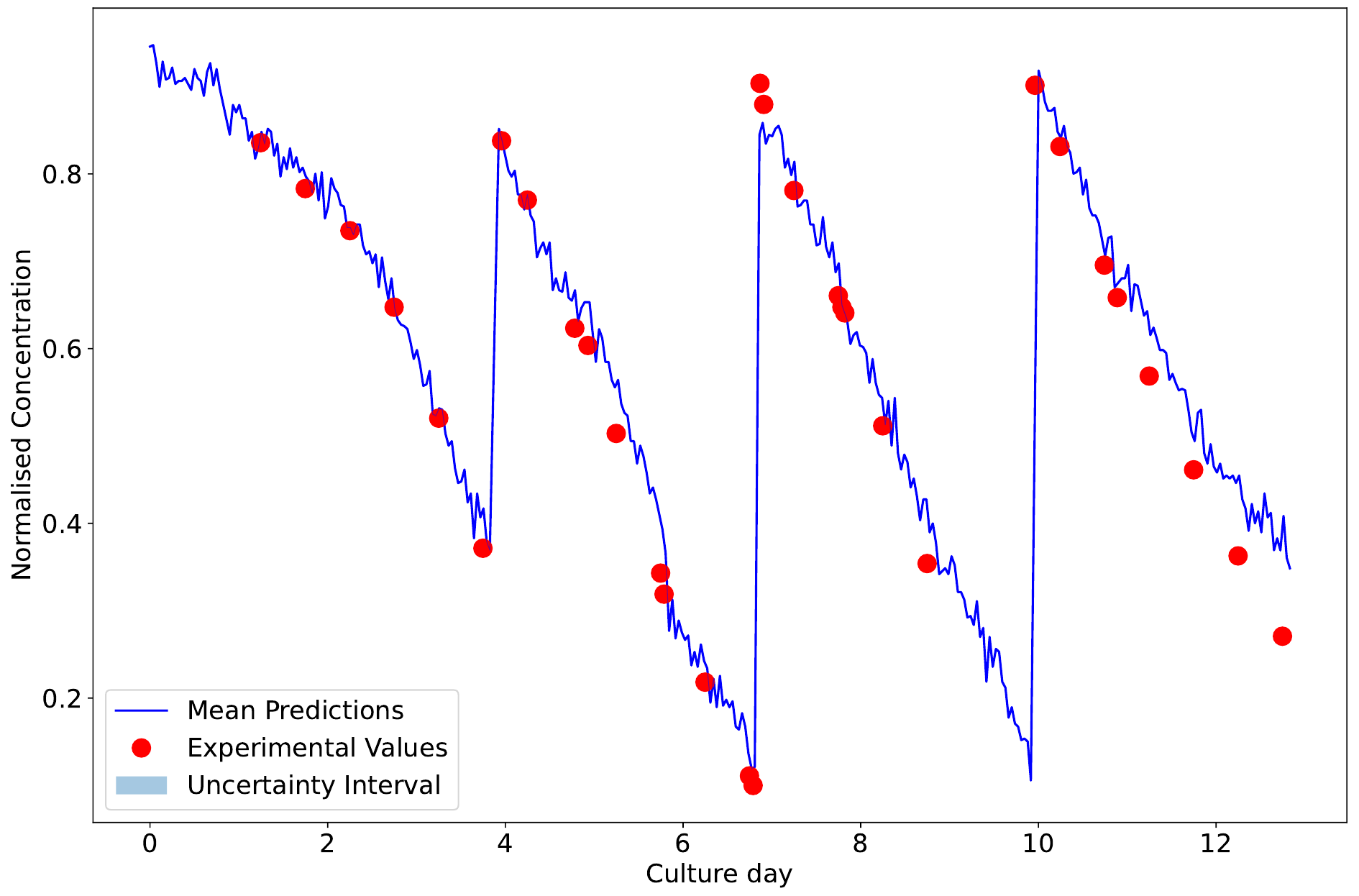}}
\end{subfloat}
\caption{Real-time monitoring of glucose concentrations in bioreactor A3 using various ML models trained on the data from bioreactor A2 and fine-tuned on the validation data from bioreactor A1.} \label{realtime_holdout}
\end{figure}

Fig. \ref{realtime_cv} and Fig. \ref{realtime_holdout} illustrate the real-time predictions of glucose concentrations using different ML models based on preprocessed Raman spectra. The demonstrations reveal that while single models can, on average, produce better predictions than the ensemble model of uncertainty-associated base learners, they fail to accurately capture the lowest glucose concentration values (bottom points) where decisions regarding glucose addition have been made or the highest glucose concentrations after glucose feeding. In contrast, the predicted values of the ensemble model, along with the standard deviation values, encompass these lowest glucose levels as well as the highest glucose concentrations immediately after glucose addition to bioreactors. These results affirm the strengths of the proposed methods in assessing the uncertainty of predicted values, which is crucial for developing control strategies for automated glucose feeding in bioreactors.

\subsubsection{Comparing the uncertainty levels of Gaussian Process models with the proposed ensemble framework}
From Figs. \ref{realtime_cv} and \ref{realtime_holdout}, it can be observed that the standard deviation of the predicted responses generated by the GP models is nearly zero. Consequently, the prediction intervals of the GP models are very narrow, limiting their ability to assess the uncertainty levels of predicted values. This limitation arises from the fact that the GP models were trained solely on offline measurements without considering the uncertainty and errors associated with the bioanalysers. In contrast, our proposed method takes into account the uncertainty associated with each offline measurement and incorporates this information during the construction of the ML models. As a result, the predicted responses, along with the corresponding standard deviation values, provide a comprehensive coverage of practical observations. The wider prediction intervals obtained using our proposed method effectively assess the uncertainty of predicted values and facilitate informed decision-making for the control process.

\begin{table}
\centering
\begin{tabular}{|m{4cm}|c|c|c|}
\hline
\multirow{2}{*}{\textbf{Coverage(\%)}} & \multicolumn{2}{c|}{\textbf{5-fold Cross-validation}} &  \textbf{Hold-out validation}   \\
\cline{2-4}
 & Bioreactor A1 & Bioreactor A3 & Bioreactor A3 \\
\hline
Ensemble of KPCA + SVRs & 100 & 100 & 100 \\
\hline
Ensemble of PLSRs & 96.97 & 96.88 & 96.88 \\
\hline
KPCA + GP & 0 & 0 & 3.125 \\
\hline
\end{tabular}
\caption{Coverage of uncertainty bounds with respect to testing samples of the proposed ensemble framework and Gaussian Process models.} \label{coverage}
\end{table}

Table \ref{coverage} shows the coverage percentage of the uncertainty regions of our proposed ensemble models and the GP models with respect to actual observations in the testing bioreactors, using 5-fold cross-validation and hold-out validation for model building and hyper-parameter tuning. It can be seen that over 95\% of the actual offline glucose concentration values fall within the uncertainty boundary of the proposed ensemble models, as expected when using $2 \cdot \sigma$ to estimate the uncertainty boundary of predictions. In contrast, the boundary of Raman-based GP models in this experiment usually does not contain the actual observations because the width of the boundary is very small. These results confirm that our proposed ensemble method estimates the uncertainty of the predictions better than the popular GP models for the real-time monitoring problem of glucose concentrations within cell culture bioreactors using Raman spectra as input features.

\section{Conclusion} \label{conclusion}
This paper introduced a novel framework capable of integrating any regressors as base learners to estimate uncertainty associated with each predictive outcome, especially in situations with limited training data. The coefficients of variation from offline measurements are utilised to calculate the standard deviation of Gaussian distributions, which, in turn, are employed to generate synthetic samples complementing the available values in the dataset. All synthetic data contribute to the training of base learners. The effectiveness of the proposed method was evaluated through two case studies. The first case involves using obtained offline measurements on the current culture day to predict mAb concentrations on the next culture day. The second case uses real-time Raman spectral data as input features to predict glucose concentrations for real-time montioring of bioreactor runs. Empirical results demonstrated the robust performance of the proposed framework in both case studies, with small testing errors. Notably, a key strength of the proposed method lies in its ability to provide the uncertainty level associated with each prediction. This uncertainty level is crucial for informed decision-making in control strategies to enhance cell culture process performance, such as adjusting glucose levels in bioreactors to sustain cell growth and productivity.

There are several potential directions for expanding the proposed framework. In the scenario where only offline measurements are used for early predictions regarding the future state of bioreactors, it becomes crucial to assess the impact of each input feature by assigning a specific coefficient of variation for each offline measurement. For online monitoring based on Raman spectral data, enhancing model accuracy could involve considering the incorporation of additional information beyond Raman spectra. This might include control variables, manual intervention data, or domain knowledge derived from computational fluid dynamics models and the kinetics of each cultivation process \cite{taki23}. Moreover, leveraging the predictive results along with the uncertainty levels provided by the proposed method could serve as a foundation for developing control strategies for real-time feedback control of bioreactors, particularly based on online glucose concentrations. In addition, there is a need to develop automated methods for combining and optimising hyperparameters of Raman data preprocessing techniques, moving beyond the fixed parameter setting used in the current work.

\bibliography{mybib}

\begin{thebibliography}{10}
\urlstyle{rm}
\expandafter\ifx\csname url\endcsname\relax
  \def\url#1{\texttt{#1}}\fi
\expandafter\ifx\csname urlprefix\endcsname\relax\def\urlprefix{URL }\fi
\expandafter\ifx\csname doiprefix\endcsname\relax\def\doiprefix{DOI: }\fi
\providecommand{\bibinfo}[2]{#2}
\providecommand{\eprint}[2][]{\url{#2}}

\bibitem{luhw20}
\bibinfo{author}{Lu, R.-M.} \emph{et~al.}
\newblock \bibinfo{journal}{\bibinfo{title}{Development of therapeutic antibodies for the treatment of diseases}}.
\newblock {\emph{\JournalTitle{Journal of Biomedical Science}}} \textbf{\bibinfo{volume}{27}}, \bibinfo{pages}{1--30} (\bibinfo{year}{2020}).

\bibitem{homo23}
\bibinfo{author}{Hong, M.~S.} \emph{et~al.}
\newblock \bibinfo{journal}{\bibinfo{title}{Smart process analytics for the end-to-end batch manufacturing of monoclonal antibodies}}.
\newblock {\emph{\JournalTitle{Computers \& Chemical Engineering}}} \textbf{\bibinfo{volume}{179}}, \bibinfo{pages}{108445} (\bibinfo{year}{2023}).

\bibitem{ph21}
\bibinfo{author}{Pharma, E.}
\newblock \bibinfo{title}{World preview 2021, outlook to 2026}.
\newblock \bibinfo{type}{Tech. Rep.}, \bibinfo{institution}{Evaluate Ltd}, \bibinfo{address}{London, UK} (\bibinfo{year}{2021}).

\bibitem{pabu19}
\bibinfo{author}{Papathanasiou, M.~M.}, \bibinfo{author}{Burnak, B.}, \bibinfo{author}{Katz, J.}, \bibinfo{author}{Shah, N.} \& \bibinfo{author}{Pistikopoulos, E.~N.}
\newblock \bibinfo{journal}{\bibinfo{title}{Assisting continuous biomanufacturing through advanced control in downstream purification}}.
\newblock {\emph{\JournalTitle{Computers and Chemical Engineering}}} \textbf{\bibinfo{volume}{125}}, \bibinfo{pages}{232--248}, \doiprefix\url{10.1016/j.compchemeng.2019.03.013} (\bibinfo{year}{2019}).

\bibitem{wu04}
\bibinfo{author}{Wurm, F.~M.}
\newblock \bibinfo{journal}{\bibinfo{title}{Production of recombinant protein therapeutics in cultivated mammalian cells}}.
\newblock {\emph{\JournalTitle{Nature Biotechnology}}} \textbf{\bibinfo{volume}{22}}, \bibinfo{pages}{1393--1398} (\bibinfo{year}{2004}).

\bibitem{paqu17}
\bibinfo{author}{Papathanasiou, M.~M.} \emph{et~al.}
\newblock \bibinfo{journal}{\bibinfo{title}{Advanced model-based control strategies for the intensification of upstream and downstream processing in {mAb} production}}.
\newblock {\emph{\JournalTitle{Biotechnology Progress}}} \textbf{\bibinfo{volume}{33}}, \bibinfo{pages}{966--988}, \doiprefix\url{10.1002/btpr.2483} (\bibinfo{year}{2017}).

\bibitem{sa22}
\bibinfo{author}{Satheka, A.~C.}
\newblock \bibinfo{title}{Upscaling of clinical grade stem cell production: Upstream processing (usp) and downstream processing (dsp) operations of cell expansion, harvesting, detachment, separation, washing and concentration steps, and the regulatory requirements}.
\newblock In \emph{\bibinfo{booktitle}{Stem Cell Production}}, \bibinfo{pages}{159--184} (\bibinfo{publisher}{Springer}, \bibinfo{year}{2022}).

\bibitem{khba24}
\bibinfo{author}{Khuat, T.~T.}, \bibinfo{author}{Bassett, R.}, \bibinfo{author}{Otte, E.}, \bibinfo{author}{Grevis-James, A.} \& \bibinfo{author}{Gabrys, B.}
\newblock \bibinfo{journal}{\bibinfo{title}{Applications of machine learning in antibody discovery, process development, manufacturing and formulation: Current trends, challenges, and opportunities}}.
\newblock {\emph{\JournalTitle{Computers \& Chemical Engineering}}} \bibinfo{pages}{108585} (\bibinfo{year}{2024}).

\bibitem{ke09}
\bibinfo{author}{Kelley, B.}
\newblock \bibinfo{journal}{\bibinfo{title}{Industrialization of mab production technology: the bioprocessing industry at a crossroads}}.
\newblock {\emph{\JournalTitle{MAbs}}} \textbf{\bibinfo{volume}{1}}, \bibinfo{pages}{443--452} (\bibinfo{year}{2009}).

\bibitem{livi10}
\bibinfo{author}{Li, F.}, \bibinfo{author}{Vijayasankaran, N.}, \bibinfo{author}{Shen, A.}, \bibinfo{author}{Kiss, R.} \& \bibinfo{author}{Amanullah, A.}
\newblock \bibinfo{journal}{\bibinfo{title}{Cell culture processes for monoclonal antibody production}}.
\newblock {\emph{\JournalTitle{MAbs}}} \textbf{\bibinfo{volume}{2}}, \bibinfo{pages}{466--479} (\bibinfo{year}{2010}).

\bibitem{deab10}
\bibinfo{author}{Derfus, G.~E.} \emph{et~al.}
\newblock \bibinfo{journal}{\bibinfo{title}{Cell culture monitoring via an auto-sampler and an integrated multi-functional off-line analyzer}}.
\newblock {\emph{\JournalTitle{Biotechnology progress}}} \textbf{\bibinfo{volume}{26}}, \bibinfo{pages}{284--292} (\bibinfo{year}{2010}).

\bibitem{mcmc12}
\bibinfo{author}{McRae, M.} \emph{et~al.}
\newblock \bibinfo{title}{Facilitating multisite bioprocess transfer: Multi-instrument and multi-platform comparability and long term stability of nova biomedical's bioprofile chemistry and gas analyzers}.
\newblock In \emph{\bibinfo{booktitle}{BioProcess International Europe}} (\bibinfo{year}{2012}).

\bibitem{hemc23}
\bibinfo{author}{Heid, E.}, \bibinfo{author}{McGill, C.~J.}, \bibinfo{author}{Vermeire, F.~H.} \& \bibinfo{author}{Green, W.~H.}
\newblock \bibinfo{journal}{\bibinfo{title}{Characterizing uncertainty in machine learning for chemistry}}.
\newblock {\emph{\JournalTitle{Journal of Chemical Information and Modeling}}} \textbf{\bibinfo{volume}{63}}, \bibinfo{pages}{4012--4029} (\bibinfo{year}{2023}).

\bibitem{huvo21}
\bibinfo{author}{Hutter, C.}, \bibinfo{author}{von Stosch, M.}, \bibinfo{author}{Cruz~Bournazou, M.~N.} \& \bibinfo{author}{Butt{\'e}, A.}
\newblock \bibinfo{journal}{\bibinfo{title}{Knowledge transfer across cell lines using hybrid gaussian process models with entity embedding vectors}}.
\newblock {\emph{\JournalTitle{Biotechnology and Bioengineering}}} \textbf{\bibinfo{volume}{118}}, \bibinfo{pages}{4389--4401} (\bibinfo{year}{2021}).

\bibitem{tuga18}
\bibinfo{author}{Tulsyan, A.}, \bibinfo{author}{Garvin, C.} \& \bibinfo{author}{Ündey, C.}
\newblock \bibinfo{journal}{\bibinfo{title}{Advances in industrial biopharmaceutical batch process monitoring: Machine-learning methods for small data problems}}.
\newblock {\emph{\JournalTitle{Biotechnology and Bioengineering}}} \textbf{\bibinfo{volume}{115}}, \bibinfo{pages}{1915--1924}, \doiprefix\url{10.1002/bit.26605} (\bibinfo{year}{2018}).

\bibitem{baal21}
\bibinfo{author}{Banner, M.} \emph{et~al.}
\newblock \bibinfo{journal}{\bibinfo{title}{A decade in review: use of data analytics within the biopharmaceutical sector}}.
\newblock {\emph{\JournalTitle{Current Opinion in Chemical Engineering}}} \textbf{\bibinfo{volume}{34}}, \bibinfo{pages}{100758}, \doiprefix\url{10.1016/j.coche.2021.100758} (\bibinfo{year}{2021}).

\bibitem{koko22}
\bibinfo{author}{Kokol, P.}, \bibinfo{author}{Kokol, M.} \& \bibinfo{author}{Zagoranski, S.}
\newblock \bibinfo{journal}{\bibinfo{title}{Machine learning on small size samples: A synthetic knowledge synthesis}}.
\newblock {\emph{\JournalTitle{Science Progress}}} \textbf{\bibinfo{volume}{105}}, \bibinfo{pages}{00368504211029777} (\bibinfo{year}{2022}).

\bibitem{gole12}
\bibinfo{author}{Goodhue, D.~L.}, \bibinfo{author}{Lewis, W.} \& \bibinfo{author}{Thompson, R.}
\newblock \bibinfo{journal}{\bibinfo{title}{Does pls have advantages for small sample size or non-normal data?}}
\newblock {\emph{\JournalTitle{MIS quarterly}}} \bibinfo{pages}{981--1001} (\bibinfo{year}{2012}).

\bibitem{xure18}
\bibinfo{author}{Xu, J.} \emph{et~al.}
\newblock \bibinfo{journal}{\bibinfo{title}{Improving titer while maintaining quality of final formulated drug substance via optimization of cho cell culture conditions in low-iron chemically defined media}}.
\newblock {\emph{\JournalTitle{MAbs}}} \textbf{\bibinfo{volume}{10}}, \bibinfo{pages}{488--499} (\bibinfo{year}{2018}).

\bibitem{kile05}
\bibinfo{author}{Kim, D.~Y.}, \bibinfo{author}{Lee, J.~C.}, \bibinfo{author}{Chang, H.~N.} \& \bibinfo{author}{Oh, D.~J.}
\newblock \bibinfo{journal}{\bibinfo{title}{Effects of supplementation of various medium components on chinese hamster ovary cell cultures producing recombinant antibody}}.
\newblock {\emph{\JournalTitle{Cytotechnology}}} \textbf{\bibinfo{volume}{47}}, \bibinfo{pages}{37--49} (\bibinfo{year}{2005}).

\bibitem{gase21}
\bibinfo{author}{Gangadharan, N.} \emph{et~al.}
\newblock \bibinfo{journal}{\bibinfo{title}{Data intelligence for process performance prediction in biologics manufacturing}}.
\newblock {\emph{\JournalTitle{Computers \& Chemical Engineering}}} \textbf{\bibinfo{volume}{146}}, \bibinfo{pages}{107226} (\bibinfo{year}{2021}).

\bibitem{peva11}
\bibinfo{author}{Pedregosa, F.} \emph{et~al.}
\newblock \bibinfo{journal}{\bibinfo{title}{Scikit-learn: Machine learning in python}}.
\newblock {\emph{\JournalTitle{Journal of machine Learning research}}} \textbf{\bibinfo{volume}{12}}, \bibinfo{pages}{2825--2830} (\bibinfo{year}{2011}).

\bibitem{aksa19}
\bibinfo{author}{Akiba, T.}, \bibinfo{author}{Sano, S.}, \bibinfo{author}{Yanase, T.}, \bibinfo{author}{Ohta, T.} \& \bibinfo{author}{Koyama, M.}
\newblock \bibinfo{title}{Optuna: A next-generation hyperparameter optimization framework}.
\newblock In \emph{\bibinfo{booktitle}{Proceedings of the 25th ACM SIGKDD international conference on knowledge discovery \& data mining}}, \bibinfo{pages}{2623--2631} (\bibinfo{year}{2019}).

\bibitem{taki23}
\bibinfo{author}{Tanemura, H.} \emph{et~al.}
\newblock \bibinfo{journal}{\bibinfo{title}{Comprehensive modeling of cell culture profile using raman spectroscopy and machine learning}}.
\newblock {\emph{\JournalTitle{Scientific Reports}}} \textbf{\bibinfo{volume}{13}}, \bibinfo{pages}{21805} (\bibinfo{year}{2023}).

\bibitem{giwa22}
\bibinfo{author}{Gillespie, C.} \emph{et~al.}
\newblock \bibinfo{journal}{\bibinfo{title}{Systematic assessment of process analytical technologies for biologics}}.
\newblock {\emph{\JournalTitle{Biotechnology and Bioengineering}}} \textbf{\bibinfo{volume}{119}}, \bibinfo{pages}{423--434} (\bibinfo{year}{2022}).

\bibitem{gima23}
\bibinfo{author}{Gibbons, L.} \emph{et~al.}
\newblock \bibinfo{journal}{\bibinfo{title}{An assessment of the impact of raman based glucose feedback control on cho cell bioreactor process development}}.
\newblock {\emph{\JournalTitle{Biotechnology Progress}}} \textbf{\bibinfo{volume}{39}}, \bibinfo{pages}{e3371} (\bibinfo{year}{2023}).

\bibitem{liko16}
\bibinfo{author}{Liland, K.~H.}, \bibinfo{author}{Kohler, A.} \& \bibinfo{author}{Afseth, N.~K.}
\newblock \bibinfo{journal}{\bibinfo{title}{Model-based pre-processing in raman spectroscopy of biological samples}}.
\newblock {\emph{\JournalTitle{Journal of Raman Spectroscopy}}} \textbf{\bibinfo{volume}{47}}, \bibinfo{pages}{643--650} (\bibinfo{year}{2016}).

\bibitem{poma22}
\bibinfo{author}{Poth, M.}, \bibinfo{author}{Magill, G.}, \bibinfo{author}{Filgertshofer, A.}, \bibinfo{author}{Popp, O.} \& \bibinfo{author}{Gro{\ss}kopf, T.}
\newblock \bibinfo{journal}{\bibinfo{title}{Extensive evaluation of machine learning models and data preprocessings for raman modeling in bioprocessing}}.
\newblock {\emph{\JournalTitle{Journal of Raman Spectroscopy}}} \textbf{\bibinfo{volume}{53}}, \bibinfo{pages}{1580--1591} (\bibinfo{year}{2022}).

\bibitem{sago64}
\bibinfo{author}{Savitzky, A.} \& \bibinfo{author}{Golay, M.~J.}
\newblock \bibinfo{journal}{\bibinfo{title}{Smoothing and differentiation of data by simplified least squares procedures.}}
\newblock {\emph{\JournalTitle{Analytical chemistry}}} \textbf{\bibinfo{volume}{36}}, \bibinfo{pages}{1627--1639} (\bibinfo{year}{1964}).

\end{thebibliography}

\section*{Acknowledgements}
This research was supported under the Australian Research Council's Industrial Transformation Research Program (ITRP) funding scheme (project number IH210100051). The ARC Digital Bioprocess Development Hub is a collaboration between The University of Melbourne, University of Technology Sydney, RMIT University, CSL Innovation Pty Ltd, Cytiva (Global Life Science Solutions Australia Pty Ltd) and Patheon Biologics Australia Pty Ltd.

\section*{Declaration of competing interest}
Robert Bassett and Ellen Otte are employees of CSL Innovation Pty Ltd. Thanh Tung Khuat and Bogdan Gabrys declare no competing interests, including no known competing financial interests or personal relationships that could have appeared to influence the work reported in this paper.

\section*{Data availability}
Data used in Section \ref{offline_performance} can be downloaded from \url{https://ars.els-cdn.com/content/image/1-s2.0-S0098135421000041-mmc3.xlsx}. However, the data used in Section \ref{monitoring} is the intellectual property of CSL Limited and is therefore not shared publicly.



\end{document}